\documentclass{article}
\usepackage{arxiv}

\usepackage[utf8]{inputenc} 
\usepackage[T1]{fontenc}   
\usepackage{hyperref}       
\usepackage{url}            
\usepackage{booktabs}       
\usepackage{amsfonts}      
\usepackage{nicefrac}       
\usepackage{microtype}      
\usepackage{lipsum}
\usepackage{graphicx}
\usepackage{amsmath}
\usepackage{txfonts}
\usepackage[dvipsnames]{xcolor}

\begin{document}

\title{Second-harmonic generation for enhancing the performance of diffractive neural networks}

\author{
    Marie Braasch\\
    Institute of Applied Physics, Abbe Center of Photonics\\
    Friedrich Schiller University Jena\\
    Albert-Einstein-Straße 15, 07745 Jena, Germany\\
    \texttt{marie.braasch@uni.jena.de}
    \And
    Anna Kartashova\\
    Institute of Applied Physics, Abbe Center of Photonics\\
    Friedrich Schiller University Jena\\
    Albert-Einstein-Straße 15, 07745 Jena, Germany\\ 
    Max Planck School of Photonics\\
    Albert-Einstein-Straße 15, 07745 Jena, Germany \\
    \And
    Sergei Krasikov\\
    Institute of Applied Physics, Abbe Center of Photonics\\
    Friedrich Schiller University Jena\\
    Albert-Einstein-Straße 15, 07745 Jena, Germany\\ 
    Heinz Nixdorf Institute, Paderborn University\\
    Fürstenallee 11, 33102 Paderborn, Germany\\
    Department of Electrical Engineering and Information Technology\\
    Paderborn University, Warburger Str. 100, 33098 Paderborn, Germany\\
    \And
    Elena Goi\\
    Institute of Applied Physics, Abbe Center of Photonics\\
    Friedrich Schiller University Jena\\
    Albert-Einstein-Straße 15, 07745 Jena, Germany\\
    \And
    Thomas Pertsch\\
    Institute of Applied Physics, Abbe Center of Photonics\\
    Friedrich Schiller University Jena\\
    Albert-Einstein-Straße 15, 07745 Jena, Germany\\
    Fraunhofer Institute for Applied Optics and Precision Engineering\\
    Albert-Einstein-Straße 7, 07745 Jena, Germany\\
    \And
    Sina Saravi\\
    Heinz Nixdorf Institute, Paderborn University\\
    Fürstenallee 11, 33102 Paderborn, Germany\\
    Department of Electrical Engineering and Information Technology\\
    Paderborn University, Warburger Str. 100, 33098 Paderborn, Germany\\
    Institute for Photonic Quantum Systems (PhoQS)\\
    Paderborn University, Warburger Str. 100, 33098 Paderborn, Germany\\
    Institute of Applied Physics, Abbe Center of Photonics\\
    Friedrich Schiller University Jena\\
    Albert-Einstein-Straße 15, 07745 Jena, Germany}

\maketitle
\begin{abstract} 
    Diffractive neural networks (DNNs) are an emerging approach for the realization of photonic artificial intelligence, especially due to their suitability for machine-vision applications and high-dimensional photonic information processing at lower power consumption. 
    However, incorporating optical nonlinear activation functions to make DNNs a feasible alternative to their electronic counterpart remains a challenge.
    Here, we investigate the inclusion of second-harmonic generation (SHG), as one of the simplest and most efficient types of optical nonlinearities, in DNNs.
    We numerically investigate the impact of SHG on the performance of classification tasks in an all-optical nonlinear DNNs. Specifically, we investigate and discuss the essential requirements for an effective arrangement of a {SHG activation} in single and multilayer DNNs. 
    We find that the performance, in terms of classification accuracy and class contrast, is affected strongly by the positioning of the {SHG activation}. Finally, we discuss and outline the constraints for including SHG in an experimental realization. Taking these constraints into account, we estimate the power-related efficiency of the nonlinear DNN system. Overall, our results provide a path towards implementing nonlinear DNNs using the SHG process.
\end{abstract}

\section{Introduction}
    Diffractive Neural Networks (DNNs) represent an optical approach to realizing neural networks, using the principles of light diffraction to perform complex computations \cite{sun2023,hu2024,chen2024}.
    At the core of a DNN lies the concept of light manipulation through structured materials. By designing diffractive optical elements, such as phase modulation layers, we can control the diffraction of light waves in a manner that mimics the operations of conventional neural networks. Dielectric metasurfaces are well suited for this task, as they provide unprecedented control over the degrees of freedom of light in diffractive settings \cite{arbabi2015}.
    \noindent DNNs offer a new approach for information processing tasks in general \cite{qian2020}, and for machine vision and image processing tasks in particular, where DNNs can either preprocess images \cite{chen2023,sakib2024} or even replace traditional neural networks with an all-optical system \cite{lin2018,goi2021,mengu2022}. 
    Such optical computing systems have the potential to offer much smaller computational latency and energy consumption compared to their electronic counterparts \cite{mcmahon2023}. 
    \\Yet, a major challenge remains: implementing an all-optical nonlinear activation function. This is because light propagation between layers is mainly a linear effect \cite{shen2006}, where optical nonlinearities are generally weak effects that manifest only at higher levels of light intensities. 
    Without nonlinearity, DNNs cannot provide a true "depth"  \cite{hornik1991}. Thus, incorporating nonlinearities is essential for realizing DNNs with complex functionalities.
    Various approaches have been explored to achieve nonlinear responses in optical neuromorphic systems \cite{destras2023,dinc2024,wang2023}, including all-optical and hybrid optoelectronic approaches, with most developments in integrated photonic platforms \cite{farmakidis2024}.
    All-optical nonlinearities are generally more challenging than hybrid ones, in terms of the needed optical power, yet could offer the full advantages of light-speed computations.
    In DNNs, only a few types of all-optical nonlinearities have been theoretically and experimentally explored, such as the photorefractive effect in crystals \cite{yan2019}, saturable absorption in thermal atomic vapors \cite{ryou2021},  electromagnetically induced transparency in laser-cooled atoms \cite{zuo2019}, combination of Kerr and two-photon absorption in a $\chi^{(3)}$ nonlinear material \cite{dong2025}, image intensifier with saturating gain \cite{wang2023}and photoluminescence from quantum dots \cite{huang2024}.
   \\These different mechanisms for incorporating nonlinearity into DNNs, each offer their own trade-offs in terms of latency, required optical power, and feasibility of a scalable implementation.
    Parametric nonlinear processes, reached in materials with $\chi^{(2)}$ and $\chi^{(3)}$ optical nonlinearities, offer the fastest type of interactions (instantaneous in practice) and the most flexibility and feasibility for implementation. Specifically, they can be tailored through micro- and nanostructuring to exhibit specific spectral and polarization responses. This can be achieved in diffractive systems using nonlinear metasurfaces \cite{zheng2023}.
    Yet parametric processes generally demand large optical intensities to manifest.
    In this category, three-wave-mixing in $\chi^{(2)}$ materials is among the most efficient types of parametric nonlinear processes. {Hence, undepleted SHG can in principle provide an instantaneous all-optical nonlinear activation, where its latency and operation speed is only limited by the speed of light propagating through the system and its optical bandwidth. Moreover, given that the spectral, spatial, and polarization properties of the SHG process can be strongly controlled through the choice of the nonlinear material and its structuring, such a nonlinear activation function is highly versatile in terms of integration in various types of technologies for realizing DNNs. Finally, the lack of a power threshold for observing the quadratic functionality in undepleted SHG, makes it robust to the total intensity of light, as long as the generated SH light can be detected feasibly.}
    It was already shown that three-wave mixing can be used to create an all-optical nonlinear activation function, similar to the ReLU function \cite{gordon2023}. Yet such an effect was manifested in a power-depleting regime, which demands a strong nonlinear interaction that is generally challenging to reach, and in this case was facilitated by the use of a nanostructured nonlinear waveguide. 
    Although reaching this regime might be possible in DNNs, it is generally more challenging compared to an integrated waveguide, as in diffractive systems we commonly deal with beams that are spread over a wider area (smaller intensities) and thinner optical elements (smaller nonlinear strengths).
    With this motivation, in our work, we investigate the use of the three-wave-mixing process of second-harmonic generation (SHG) in the undepleted regime as the optical activation layer in a DNN. Undepleted SHG creates a quadratic functionality between the input fundamental harmonic electric field (at frequency $\omega$) and the generated second-harmonic electric field (at frequency $2\omega$) \cite{boyd2008}. We are motivated by the fact that undepleted SHG is much more feasible to implement than a depleted interaction, and, in fact, there is no required nonlinear efficiency threshold for a photonic structure with $\chi^{(2)}$ nonlinearity to observe the quadratic functionality of SHG.
    We point out that the use of SHG in the undepleted regime has recently been explored in a type of reservoir computing diffractive system, made of disordered nonlinear nano particles, and showed enhancement in the performance of the system in computational accuracy \cite{wang2024}.\\
    In our work, we theoretically treat the problem of image classification using DNNs and explore the use of a {SHG crystal} to enhance the performance of such DNN systems.
    We will discuss the requirements for a practical setup, explore their potential, and finally present numerical results on their performance for the MNIST digit, the MNIST Fashion datasets, and the EMNIST handwritten letters with 10, 10, and 47 classes, respectively \cite{lecunn, xiao2017fashion, cohen2017}.
    We identify that SHG can in fact make a meaningful increase in the performance of a DNN for image classification, both in terms of classification accuracy and class contrast. But we also observe that these improvements are highly dependent on the position of the {SHG activation} in the DNN system, and there are scenarios in which the {SHG activation} can in fact worsen the performance of the system.
\section{DNN Architecture and Training}
    \subsection{DNN Parameters}
    \begin{figure}[htbp]
        \centering
        \includegraphics[width=1\linewidth]{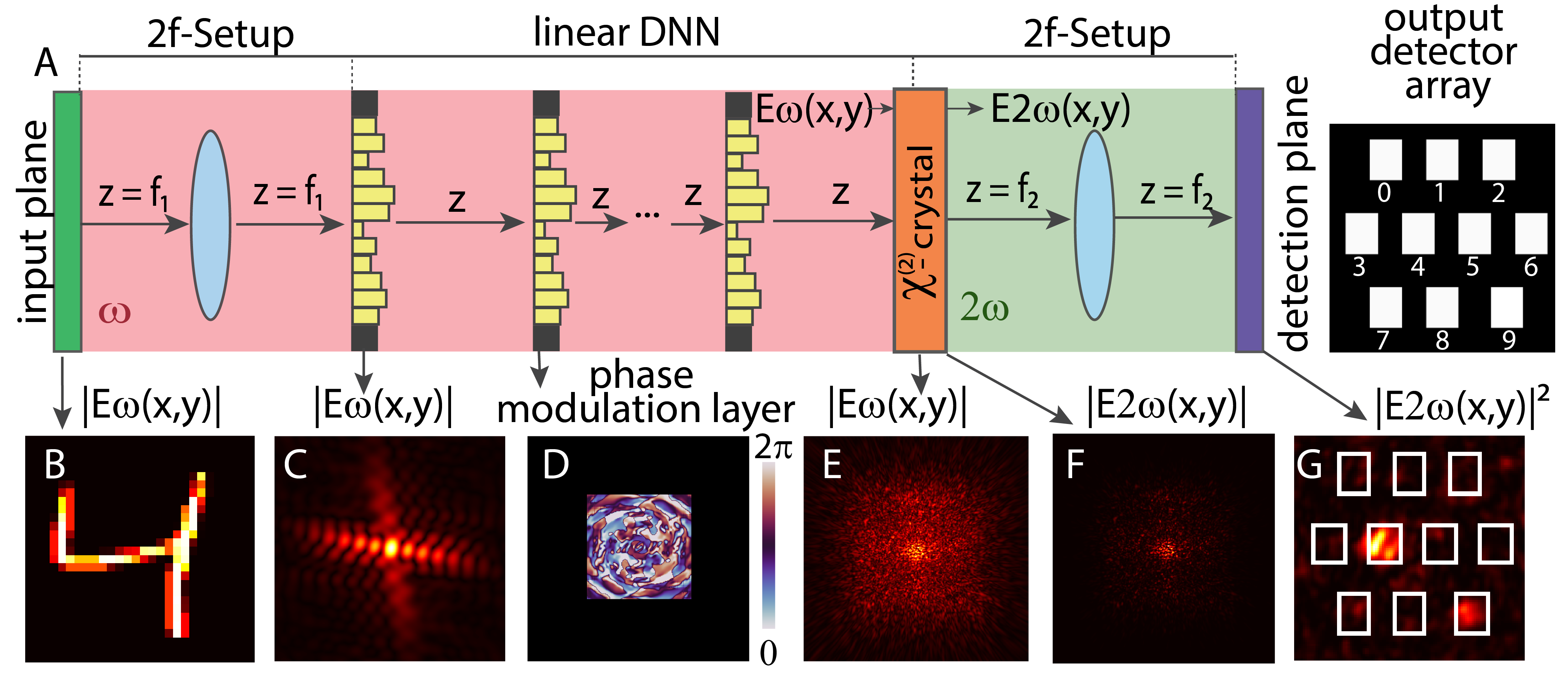}
        \caption{(A) The nonlinear DNN consists of an amplitude encoded input, followed by a 2f-Setup,  a single or a series of phase modulation masks, a \(\chi^{(2)}\)-crystal and a second 2f-Setup.  Light around the phase modulation layers is blocked, such that only lower spatial frequency components are transmitted. The DNN is trained to focus light into specific regions at the output plane, corresponding to different classes as illustrated by the output detector array. Below the setup, we plot the corresponding amplitude at the input plane (B), after the first 2f-setup, delivering a Fourier transform (C), after an exemplary phase modulation layer (D), the amplitude before (E) and after (F) the \(\chi^{(2)}\)-crystal, and the resulting intensity at the detection plane (G).}
        \label{fig:model}
    \end{figure}
    In this work, we consider a Fourier-space DNN ~\cite{yan2019}, where the input images (see Fig.~\ref{fig:model}(A) are amplitude-encoded (B) and pass through a 2f-lens system, which performs a Fourier transform (see Fig.~\ref{fig:model}) before projecting the input onto the first layer of a DNN (C). After propagation through the DNN, another 2f-system projects the Fourier transform of the output plane of the DNN to the detection plane.
    The DNN consists of a single or a series of cascaded phase-modulation layers with free-space propagation between them. The phase modulation layers (see Fig.~\ref{fig:model}(D)) add a spatially-varying phase-shift to the complex field.
    The phase shift at each spatial point of each phase modulation layer determines a multiplicative bias term in this optical neural network (to be found through training), while the weights are determined by the free-space response function in the Rayleigh-Sommerfield diffraction formula (see supplementary, Eq.~S2). We train our DNN to focus light into distinct regions of the detection plane, each region representing a different class, as illustrated by the ideal output in Fig.~\ref{fig:model}. Then, the class is determined by the region where the light intensity is highest, see Fig.~\ref{fig:model}(G).
    The system also includes a nonlinear {SHG activation}, schematically modeled here by a $\chi^{(2)}$-crystal.
    The role of the {SHG crystal} is to apply a pointwise nonlinear transformation to the complex field, i.e. \(E_{2\omega}(x,y) = E_\omega(x,y)^2\) (see Fig.~\ref{fig:model}(E) and Fig.~\ref{fig:model}(F). In this way, the field-squaring SHG response introduces a physical nonlinearity into this optical neural network.
    In the following sections, we compare configurations without SHG or including SHG at different positions in a multi- or single-layer DNN.
    Note that in configurations with SHG, the DNN performs its operation on light of frequency $\omega$ before the {SHG crystal}, and after it all operations are performed on frequency-doubled light at $2\omega$. That also means that the $\omega$ field does not contribute to the detected signal; in an experimental setting, this can be achieved by introducing spectral filters, since the \(\omega\) and \(2\omega\) are well separated. 
    \subsection{Numerical modeling and training} 
    To model free-space propagation, we followed the numerical method introduced by~\cite{shen2006} (see supplementary section 1 and 2).
    
    We set a fixed number of neurons (phase-modulation pixels) of 56\(\times\)56 for each linear DNN phase modulation layer modeled with thin element approximation. Any light around the phase modulation layer is blocked. 
    We trained the phase modulation layers using the Adam optimizer in Keras~\cite{keras}, with the sparse categorical cross-entropy loss function (see supplementary section 3).
    It should be emphasized that the parameters for each configuration are individually optimized using a Bayesian optimizer to have a fair comparison \cite{BO, stander2002}.
    The optimized parameters include the size of the magnification of the Fourier plane as well as the magnification of the output plane.
    Further details on the optimization process, as well as the corresponding physical parameters, are explained and documented in the supplementary section 4 and 5.
    In any configuration including free-space propagation, either in between layers or after the last layer, we fix the distances to 30\(\lambda\). This choice is made to maximize the expression capabilities of the DNN \cite{Zheng:22}.
    
\section{Results}
\subsection{SHG placement in Single-Layer DNNs}

\begin{figure}[htbp]
        \centering
        \includegraphics[width=0.88\linewidth]{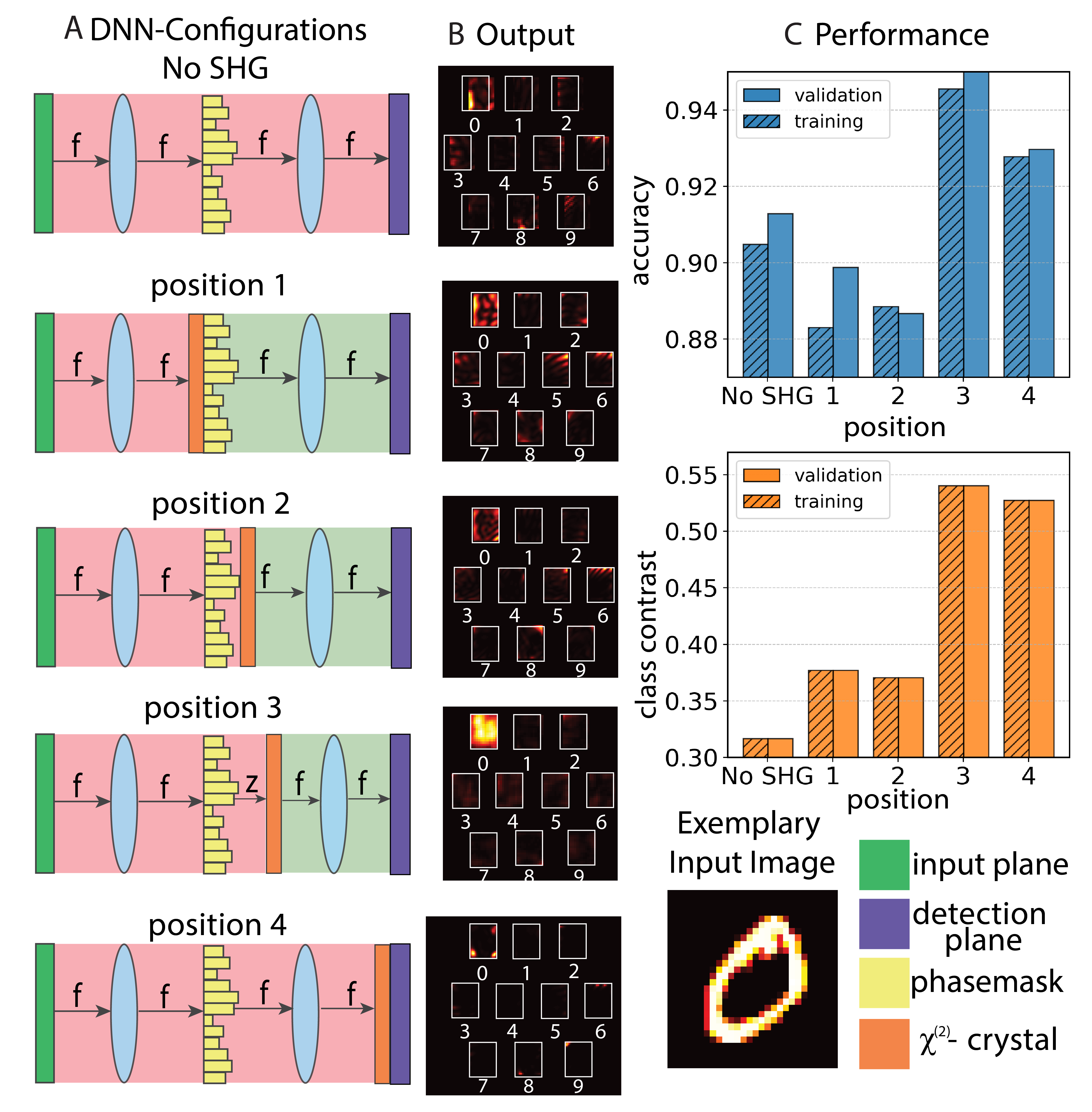}
        \caption{{SHG activation} positioned at different positions in a single-layer DNN. The figure displays the different DNN configurations (A), the intensity at the detection plane for a sample input digit (B) and the corresponding accuracy and class contrast for the  MNIST digit dataset after training (C). }
        \label{fig:Setups-SingleLayer}
    \end{figure}

    In the following, we demonstrate through our simulations that introducing SHG as a nonlinearity into a DNN can be beneficial, and we discuss the conditions required for it to be effective. We found that optimal placement of the {SHG activation} within the DNN is essential to achieve the desired performance improvement.
    To gain some general understanding, we start with a simple DNN; in particular, we consider a DNN with a single linear layer, which is realized by a phase-modulation mask. We also want to highlight the practical relevance of these single-layer DNNs, as they are easier to implement in real-world applications due to the absence of alignment in-between the phase modulation layers. Although here we extract our intuitions based on a single-layer DNN, we show later that this understanding also holds in a multilayer DNN.
    In a single-layer DNN, the {SHG activation} can be positioned directly before or after the phase modulation layer, with some propagation $z_{SHG}$ after the phase modulation layer, or right before the detector. The different configurations are summarized schematically in Fig.~\ref{fig:Setups-SingleLayer}(A). For reference, the figure includes a configuration without SHG, corresponding to the fully linear DNN. 
    We optimize each DNN configuration for the classification of the MNIST digit dataset, where each input image is encoded into a binary intensity pattern. 
    To assess the classification accuracy, the intensity \(I\) is integrated within a predetermined area specific to the class in the detection plane, and the predicted class is assigned based on the patch with the highest integrated intensity. Similarly, the loss during training is determined based on the integrated intensities, as described in the supplementary materials section~3. The contrast \(C_i\) for class $i$ is defined by the ratio of the integrated intensity focused in the correct class, \(I_{i}\) and the intensity in all other $N$ class areas \(C_i=\frac{I_{i}}{\sum^N_n I_n}\). This value is averaged over the entire dataset. Note that incorrectly classified samples are excluded from this calculation to avoid bias introduced by systems with lower accuracy and a higher number of misclassified examples.
    These results are summarized in Fig.~\ref{fig:Setups-SingleLayer}(C). The optimization parameters for each configuration are detailed in the table~S1 in the supplementary materials section~S5. 
    \paragraph{Effect of SHG position on accuracy}\mbox{}\\
    We first analyze the validation accuracy.  As shown by the figure including a {SHG activation} can either increase or decrease classification accuracy depending on its position. The validation accuracy is highest for position 3, i.e. the configuration where we have some propagation distance in between the phase modulation layer and the {SHG activation} ($\chi^{(2)}$-crystal). Compared to a DNN without SHG there is an improvement of around 4\% in classification accuracy (91.3\% to 95.2\%).
    In particular, it also performs better than the configuration in which the {SHG activation} is placed at the detection plane. SHG at the detection plane essentially adds to the nonlinearity that is already present. During readout, the detector measures the sum over spatial pixels of $|E^2|$, with the addition of SHG at the detection plane, this becomes the sum over $|E^4|$. This configuration corresponds to a single final nonlinear layer with no further optical propagation, and therefore does not contribute to network depth.
    In contrast, embedding SHG within the optical network allows the nonlinearity to act on the optical field, enabling potential multiple successive nonlinear transformations and propagation stages. This intermediate nonlinearity is necessary to realize true network depth, while also preserving both amplitude and phase for a more expressive computation.
    Another important observation is, that placing SHG at position 3 performs significantly better than position 1 and 2, where the SHG is placed directly at the phase modulation layer. The positioning at 1 and 2 in fact perform even worse in validation accuracy compared to the fully linear configuration without SHG.
    Let us first focus on position 1.
    Here, the Fourier transform of the input is first squared by the {SHG activation} and is then input into the phase modulation layer. This could be thought of as a nonlinear transformation of the input, with the DNN starting at the first phase modulation layer.
    In the context of neural networks, a transformation or nonlinear mapping of the input can be advantageous, as studied in the framework of reservoir networks \cite{fischer2023}. However, it can also lead to a disadvantageous mapping \cite{saeed2025nonlinear}.
    In our DNN, the squaring of the input field, here a Fourier transform, essentially enhances the contribution of the lower spatial frequency components in the input, which are already more dominant (see Fig.~\ref{fig:model}(b)). This limits the contribution of the higher spatial frequency components that carry more information about the input's shape.
    We note that this behavior might depend on the dataset choice, e.g. in cases where the lower spatial frequencies carry more desired information or higher spatial frequencies are undesired noise. Then, suppressing higher spatial frequencies by squaring might be beneficial.
    
\begin{figure}[htbp]
        \includegraphics[width=0.9\linewidth]{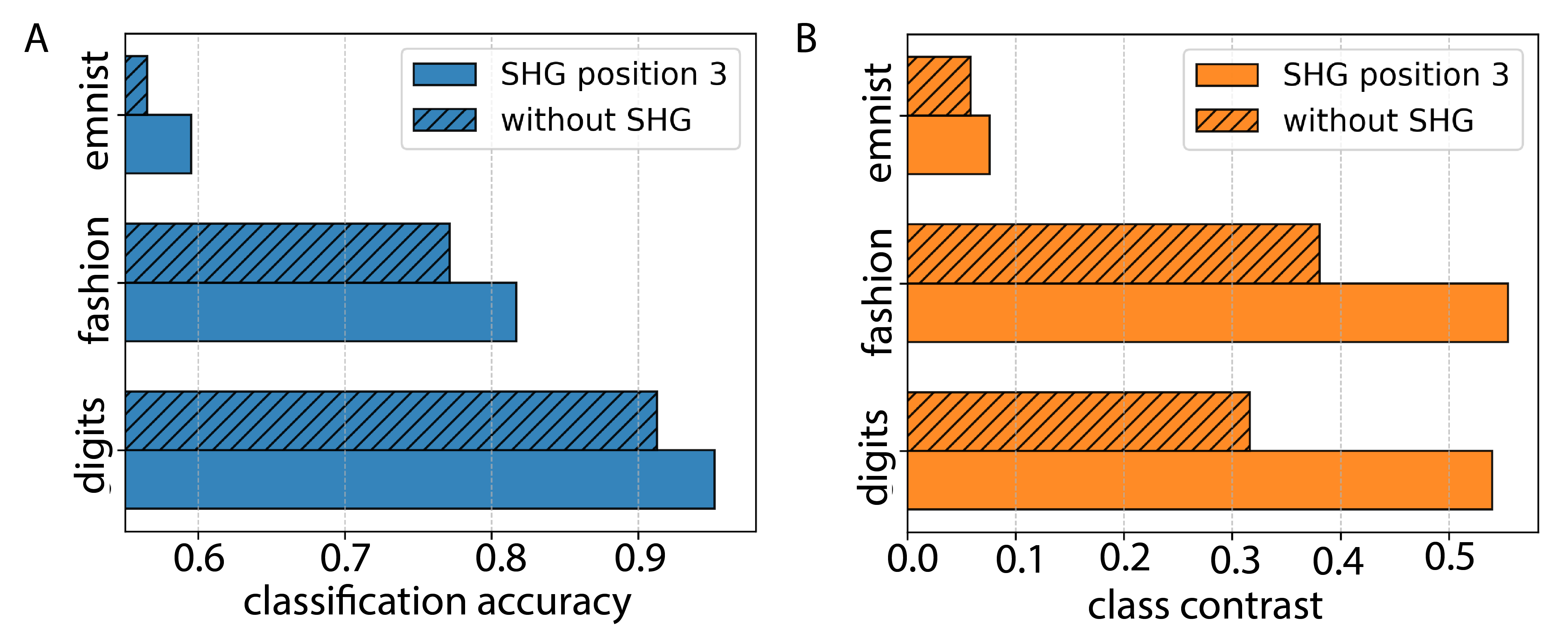}
        \centering
        \caption{(A) Accuracy and (B) class contrast for a single-layer DNN. We compare the performance of the DNN without SHG and SHG at position 3 across the MNIST digit, fashion MNIST, and EMNIST handwritten letters datasets.}
        \label{fig.Datasets}
    \end{figure}

    We can now focus on position 2. Here, we can use the same reasoning for position 1, by noticing that the two configurations have a very similar mathematical functionality: For position 1, we first square the Fourier transform and then apply the phase modulation to the field ($\left[E_\mathrm{FT}(x,y)\right]^2 e^{i\phi_1(x,y)}$), whereas for position 2, we first apply the phase modulation to the Fourier transform and then square the field ($\left[E_\mathrm{FT}(x,y)e^{i\phi_2(x,y)}\right]^2$). Hence, with the relation $2\phi_2(x,y)=\phi_1(x,y)$ between the phase profile of the phase modulation layers, the two configurations can have the exact same mathematical functionality.
    This implies, that positioning the {SHG activation} immediately after the phase-modulation layer can likewise be conceptualized as a neural network with nonlinearly transformed input.
    Hence, a comparable performance is expected for positioning SHG at positions 1 and 2, which is in fact the case, as can be seen in Fig.~\ref{fig:Setups-SingleLayer}(C).    
    The small difference in classification accuracy between both configurations can be contributed to fluctuations in validation accuracy of about 1$\%$ (see Fig.~S4 in supplementary materials section~S6), which is caused by statistical effects such as training data shuffling and different convergence behavior due to a gradient difference at the phase modulation layer.
    \paragraph{Effect of SHG position on class contrast}\mbox{}\\
    When comparing the class contrast of the different configurations, we observe an improvement for all configurations that include SHG, regardless of the layer position.
    This is also apparent upon inspection of Fig.~\ref{fig:Setups-SingleLayer}(B), where we display the intensity distribution at the detection plane for an example input digit for the different configurations. In the representation the output field is multiplied with the detection mask, i.e. the light outside the patches is not illustrated.
    The enhancement in class contrast is most visible for positioning SHG at position 3.  
    Here we achieve a class contrast of 0.54, that is, 54\% of the light within the 10 patches is focused into the desired patch on average, compared to only 31\% for the DNN without SHG.
    This could be a useful property for facilitating classification in practice, potentially making the DNN more robust to unwanted sources of noise, such as background light.
    We note, that it is already known that the class contrast can strongly depend on the loss function used in the training process, and in general there seems to be a trade-off between accuracy and class contrast performance in DNNs, which can be traversed by the choice of different loss functions~\cite{mengu2020}. Hence, we want to highlight that by use of the {SHG activation}, we are able to improve the accuracy and class contrast simultaneously bypassing the trade-off. 
    \paragraph{Performance of SHG on different datasets}\mbox{}\\
    Lastly, we repeated the analysis for the MNIST fashion and EMNIST handwritten letters dataset, to confirm that our observations are generally applicable to different problems. In Fig.~\ref{fig.Datasets} we show the improvement in class contrast and accuracy for these different datasets, where we compare the performance of SHG at position 3 and the DNN without SHG. In all cases, we observe a consistent improvement in validation accuracy of  77.1\% to 81.6\% for the MNIST fashion dataset and 56.5\% to 59.5\% for the EMNIST handwritten letter datasets. The class contrast also shows a consistent improvement of    38.1\% to 55.4\% for the MNIST-fashion and 5.9\% to 7.6\% for the EMNIST dataset.
    {
    The magnitude of improvement introduced by the SHG crystal depends on the complexity and nature of the classification task. In the case of EMNIST, the large number of classes  and the inherent similarity between handwritten characters likely limit the gains achievable with a single nonlinear stage. Several strategies could further enhance performance. One strategy could be cascading multiple nonlinear layers to increase the expressive power of the optical system. Yet this approach could be limited in practical scalability, both due to the efficiency of the undepleted SHG process and also due to the change in the frequency of light in each consecutive nonlinear layer. An approach with more potential for scalability could be based on combining SHG and parametric amplification of the fundamental harmonic, as done in Ref.~\cite{gordon2023} with a nonlinear waveguide. To realize this however, strong nonlinear interactions in the depleted regime are needed in a diffractive system. Nonetheless, how such a depleted nonlinear effect could enhance the performance of DNNs remains to be explored. Rearranging and reshaping the classification patches could also be a way of improving the performance \cite{kaden2026}. Importantly, even the modest improvements observed here are noteworthy, as the addition of the SHG crystal simultaneously enhanced both classification accuracy and class contrast,  two metrics that typically compete with one another ~\cite{fan2025joint}.}
    
\subsection{SHG placement in Multi-Layer DNNs}
\begin{figure}[htbp]
    \centering
        \includegraphics[width=1\linewidth]{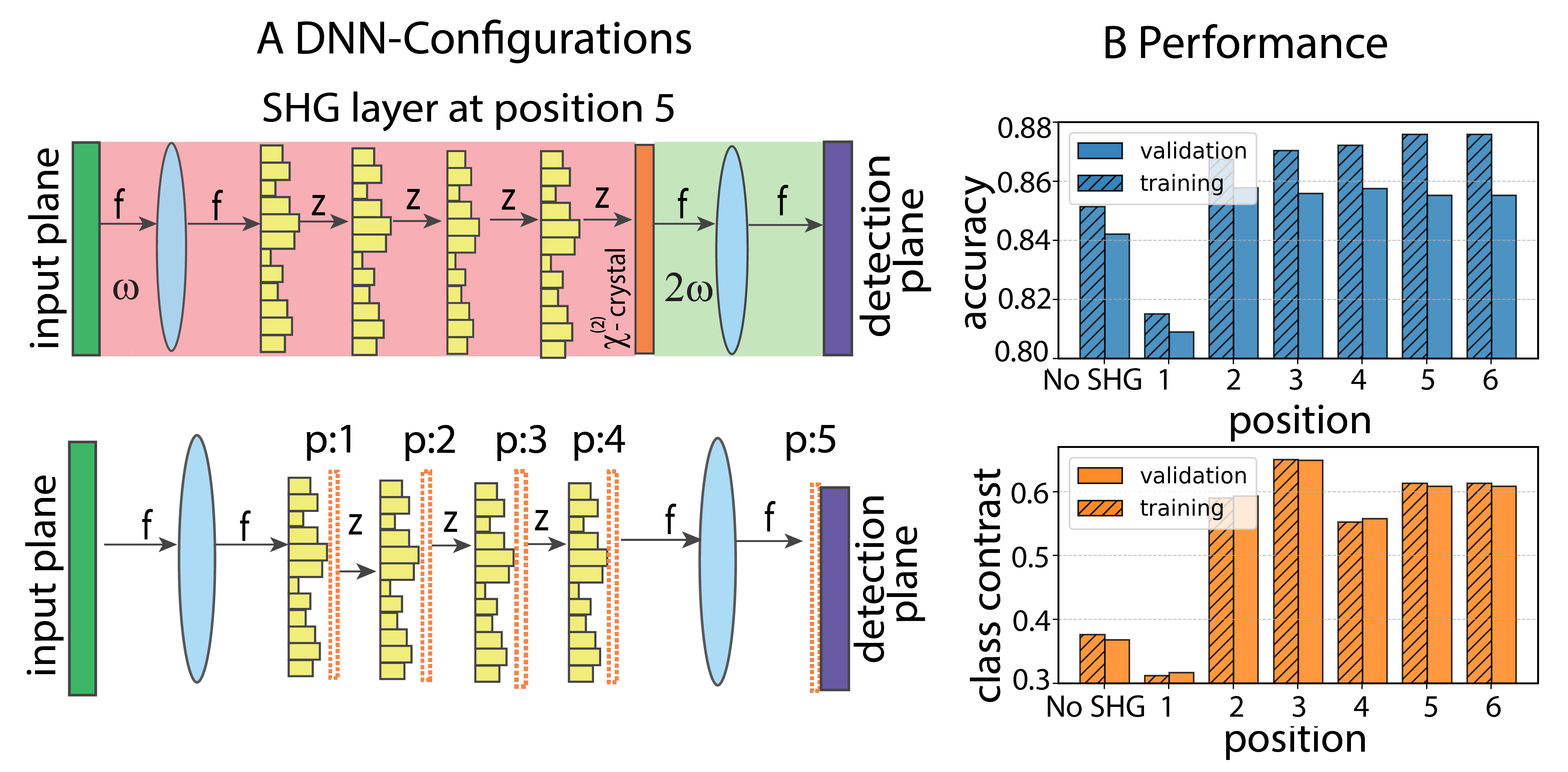}
        \caption{Schematic of the four-layer DNN with the {SHG crystal} located at different positions (A), and its performance in terms of accuracy and class contrast at the detection plane (B)}
        \label{fig:MultiLayer}       
    \end{figure}

    From the previous results, we saw that some propagation after the phase modulation layer is required before introducing SHG to achieve the best improvement in performance, in both validation accuracy and class contrast.
    In this section, we check this result in a DNN with multiple phase modulation layers, to see how this result can be generalized.
    Here we focus on a 4-layer phase modulation layer DNN, as shown in Fig.~\ref{fig:MultiLayer}(A), where the different positions of the {SHG activation} are also shown. 
    Note that the DNN performs its operation on light of frequency \(\omega\) before the {SHG crystal}, and on frequency-doubled light at 2\(\omega\) after it. In particular that means that the phase-modulation layers must be designed and modeled to provide the intended phase response at their respective operating frequencies
    The corresponding results in terms of validation accuracy and class contrast are summarized in Fig.~\ref{fig:MultiLayer}(B). Here we present the performance of the multilayer DNN for the MNIST Fashion dataset.
    The fashion MNIST is a more complex dataset compared to digit MNIST, and it allows us to check our results obtained in the previous section regarding placement dependence, not only with a different number of layers, but also using a different dataset, to show the general nature of our results.
    Details concerning the optimization parameters are summarized in supplementary materials section~4 and section~5.
    As in the section before, we use 56\(\times\)56 neurons per layer.  
    The layer distances, as well as the propagation distance to the {SHG activation}, are set to 30\(\lambda\) as before.
    Fig.~\ref{fig:MultiLayer}(B) shows generally similar trends as in the single-layer DNN. Just as before, if the {SHG crystal} is placed right after the first phase modulation layer, the accuracy is worse compared to a DNN without SHG. A difference compared to the single-layer DNN is, that here the class contrast and accuracy decreases for positioning SHG at position 1, i.e right at the first phase modulation layer. For any other SHG position, the classification accuracy and class contrast improve compared to the DNN without SHG. Hence, based on both sets of single-layer and multi-layer simulations, one can conclude more generally that the class contrast and accuracy improve both when the {SHG crystal} is at a propagation distance to the first phase modulation layer, and placing the {SHG crystal} right before or after the first phase modulation layer could degrade the accuracy and sometimes the class contrast compared to the DNN without SHG.    
    In terms of validation accuracy, the SHG configurations at positions 2-6 perform similarly to each other, with differences in the range of natural fluctuations during training (see supplementary materials section~6). In terms of class contrast, positioning SHG after at position 3, outperforms the other DNN configurations. However, any positioning within the DNN, that means after a phase modulation layer and some propagation distance leads to a consistent improvement in class contrast compared to the DNN without SHG or the DNN with SHG at position 1, directly at the first  phase modulation layer.
    When comparing the performance of the SHG at position 5 with the DNN without SHG, the validation accuracy increases from 84.2\% to 85.7\% and the class contrast from 38.1\% to 60.5\%.
    Although it is hard to generalize which DNN configuration performs best, we can see that the configurations where SHG is placed with some propagation distance after the last phase modulation layer (position 5 in the multi-layer DNN and position 3 in the single-layer DNN), lead to a consistent improvement in terms of both classification accuracy and class contrast. 
    This is a favorable outcome for practical implementations, as it allows for the fabrication of a single monolithic multi-layer linear DNNs (e.g., made from 3D nano-printing \cite{goi2021,goi2022}) that is followed by a nonlinear $\chi^{(2)}$-crystal, rather than trying to have two well-aligned separate DNNs on two sides of a $\chi^{(2)}$-crystal.
    
\section{Path towards a realistic implementation}
    In this section, we propose a path for realistic implementation given the DNN as defined above.
    In doing so, we investigate the possible physical and practical constraints regarding this implementation and provide an estimate of the achievable output power within the system for a given input power. To assess how well the DNN concentrates optical power into the correct class, three key factors must be considered.
    Firstly, there is the  conversion efficiency  of the SHG-process.
    Additionally, the phase modulation layer introduces aperture-related losses.
    Finally, there is the intrinsic focusing efficiency \(\epsilon\), describing how effectively power can be directed toward the correct output class. 
    Together, these effects define the usable fraction of optical power available for classification.
    We point out, that our use of the nonlinear crystal follows the same principle as the widely adopted use of \(\chi^{(2)}\)-crystals for parametric up-conversion imaging  ~\cite{barh2019parametric, qiu2018spiral}.  
    Here, we derive approximations, to demonstrate the practicality of our proposal regarding the use of SHG to support the assumptions made we backup the approximation with rigorous simulations.
    {
    Note, that here we assume that SHG is realized in an xy-cut potassium titanyl phosphat (KTP)-crystal 
    using Type-II phase matching, for frequency doubling from \(\lambda\)=1064 nm to \(\lambda\)=532 nm.
    The crystal is oriented such that the propagation direction lies in the xy-plane, forming an angle \(\phi = 23.5^\circ\)
    with the crystal x-axis and \(\theta = 90^\circ\) with respect to the z-axis. This configuration results in an effective nonlinear coefficient of \(d_{\mathrm{eff}}\)= 3.59 pm/V \cite{EKSMA2022KTP, boyd2008}. To maximize the conversion efficiency, the input polarization is set at \(45^\circ\) relative to the crystal z-axis. This orientation provides equal electric-field components parallel and perpendicular to the z-axis, which are required for the Type-II nonlinear interaction. The resulting second-harmonic field is polarized perpendicular to the crystal z-axis.}
    \subsection{SHG conversion efficiency under the condition of negligible diffraction}
        In order to relate the simulated fields to a physically realizable nonlinear DNN, we now examine the constraints and the expected conversion efficiency under realistic operating conditions. 
        \par
\begin{figure}[htbp]
            \centering
            \includegraphics[width=0.5\linewidth]{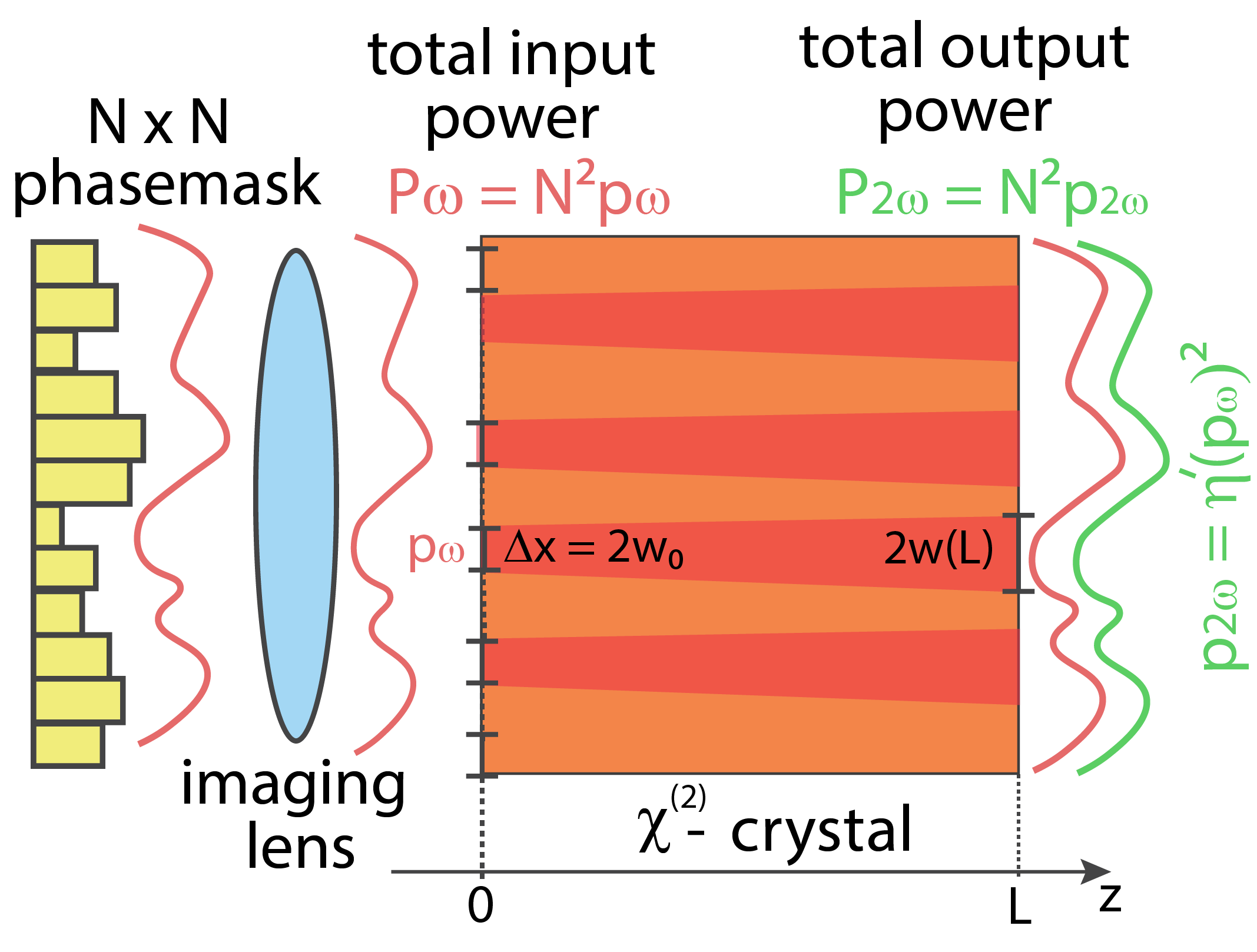}
            \caption{
            Schematic showing the envisioned use of a \(\chi^{(2)}\) bulk crystal in our proposed nonlinear DNN. At the left, we have the final layer of the linear DNN. 
            In practice, placing the nonlinear crystal right after or after some propagation distance after the final phase modulation layer can be realized by imaging the fundamental harmonic (FH) output from the final linear layer directly, or after some propagation distance \(z\) onto the input facet of the nonlinear crystal. This also allows for controlling the overall size of the FH beam, and consequently the transverse minimum feature size \(\Delta x\), at the input facet of the \(\chi^{(2)}\)-crystal.}
            \label{fig:power_schematic}
        \end{figure}

 As a rough approximation, we imagine that the field incident on the crystal can be discretized as a collection of Gaussian beams as shown in Fig.~\ref{fig:power_schematic}. 
        In this approximation, the minimum transverse spatial feature sizes in the incoming beam, \(\Delta x\), are related to the beam waist radius of the Gaussian beam \(w_0\), by $\Delta x=2w_0$. The Rayleigh length \(z_R\)  of each beam can be defined as \cite{saleh2019fundamentals}
        \begin{flalign}\label{eq:zR}
            z_R = \frac{\pi w_0^2}{\lambda_{\text{eff}} }= \frac{A}{\lambda_{\text{eff}} },
        \end{flalign}
        with \(\lambda_{\text{eff}} = \frac{\lambda}{n}\), where \(n\) is the refractive index within the crystal. The Rayleigh length describes the propagation distance where \(w(z_R) = \sqrt{2}w_0\).
       \par In the context of a set of Gaussian beams, the condition that there is no transverse mixing means, that the transverse profiles of each beam changes very little during propagation: \(w(z=0) \approx w(z=L)\approx w_0\). This corresponds to the condition that the Rayleigh range $z_R>>L$. That means, the crystal must be short enough for diffraction to be negligible. 
       \par Simultaneously, we want to generate a measurable amount of SH. That means the bulk crystal should not be too short.
       This creates a trade-off for a practical realization: We need the crystal to be longer for efficient SHG but short enough to have negligible diffraction inside the crystal.
       To better understand this trade-off and how it affects the resulting SH power, we analyze the contribution of each Gaussian beam separately. 
       The conversion efficiency \(\eta'\) relates the input and output powers of a single beam via \(p_{2\omega} = \eta' p_{\omega}^2\). According to ref.\cite{barh2019parametric, boyd2008} for a  phase matched crystal and constant beam area \(A_{in} = A_{out} = A\), (here for a single beam \(A = \pi(\frac{\Delta x}{2})^2\)), it can be derived that
        \begin{flalign}\label{eq:Power}
            \eta' = \frac{2 \omega^2 d_{eff}^2 L^2}{n_{2\omega} n_\omega^2 c^3 \epsilon_0 A },
        \end{flalign}
         where  $d_{\rm eff}$ is the effective nonlinear coefficient, $\omega$ is the fundamental harmonic frequency, $n_{\omega}$ and $n_{2\omega}$ are the refractive indices at the fundamental- and second-harmonic frequencies, $c$ is the speed of light in vacuum, and $\epsilon_0$ is the vacuum permittivity. 
         We now substitute the feature size \(\Delta x^2\) into the effective area of the conversion efficiency eq.~\ref{eq:Power} under the constraint 
         \begin{flalign}\label{eq:approximationQuality}
         z_R = bL
         \end{flalign}using  eq.~\ref{eq:zR}, where factor $b$ here indicates how much larger the Rayleigh length is with respect to the crystal length.
         Then we find
         \begin{flalign}\label{eq:linear}
             \eta\propto\frac{L^2}{A}\propto\frac{L^2}{\Delta x^2}\propto\frac{L}{b\lambda_{\rm eff}}.
         \end{flalign}
         As we can see, an intrinsic trade-off arises from our constraint, increasing \(b\) suppresses transverse mixing, but simultaneously reduces the conversion efficiency. Interestingly, while conversion efficiency typically scales quadratically  with the crystal length, under the diffraction constraint it increases only linearly. 
        \subsection{Validity of SHG propagation model}
        
\begin{figure}[htbp!]
            \centering
            \includegraphics[width=1\linewidth]{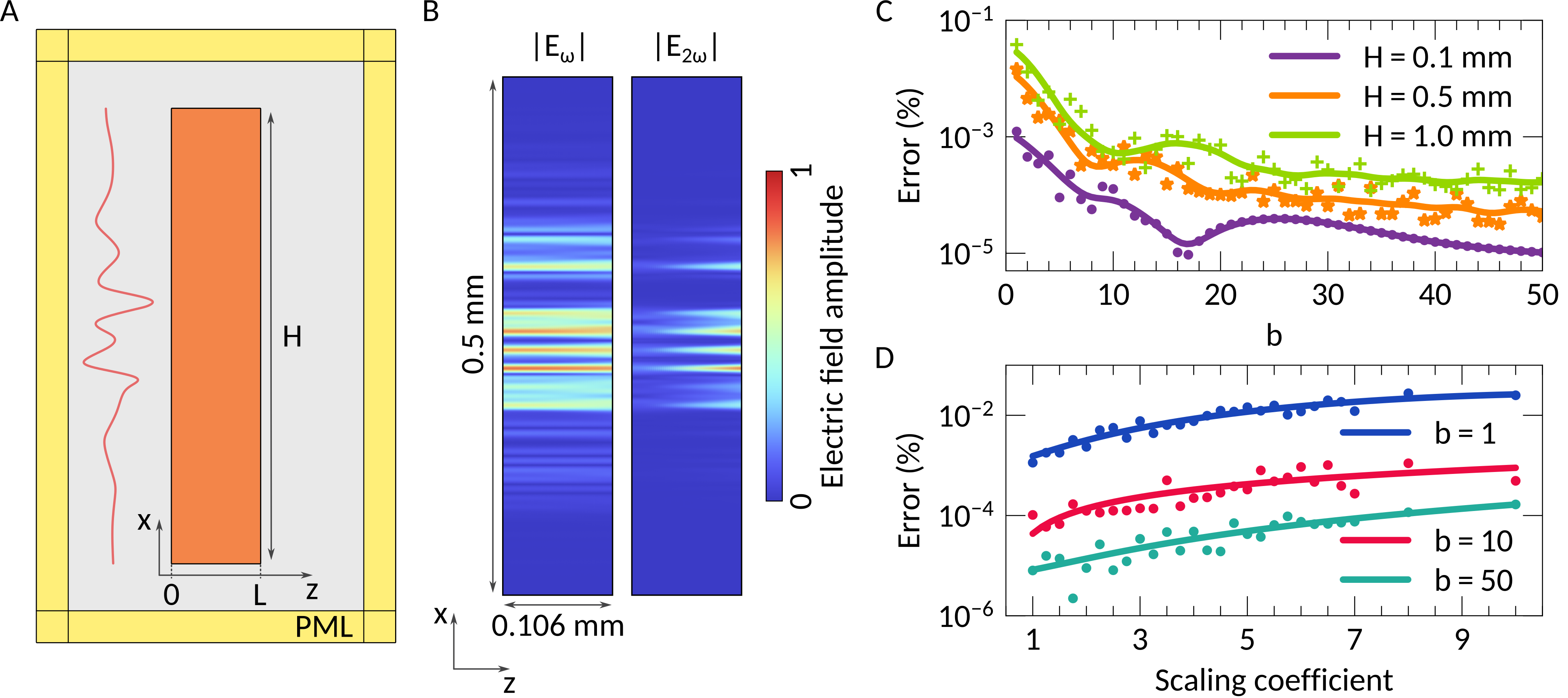}
            \caption{{(A) Schematics of the 2D full-wave nonlinear simulation model consisting of a nonlinear crystal in a free space background. The length of the crystal is $L$ and its transverse size if $H = d_{\text{tot}}$, where $d_{\text{tot}}$ is the total beam diameter. The beam contains $N = 56$ spatial features with randomly defined amplitudes. (B) Example of the normalized field distributions found from the full-wave simulation at the excited fundamental- and the generated second-harmonic frequencies, for the specific case of $H=0.5$ mm and $L=0.106$ mm, corresponding to $b=1$. Only the region inside the crystal is displayed here. (C) Proportionality error between the squared fundamental-harmonic field before the crystal $E_{\omega}^2(z=0)$ and the SH field after the crystal $E_{2\omega}(z=L)$, as a function of the $b$ coefficient that quantifies the extent of the negligible-diffraction approximation. Results are calculated for 3 different values of the transverse size $H$. Solid lines on the plot are obtained as smoothing spline fit lines, for better visualizing the data dependencies. (D)~Proportionality error calculated as a function of the transverse size $H$, for 3 different values of the $b$ coefficient. The transverse size is expressed through a scaling coefficient, with $H=\mathrm{scaling\: coefficient}\times H_\mathrm{ref}$ and $H_\mathrm{ref} = 0.1$ mm. Solid lines on the plot are obtained as polynomial fit lines, for better visualizing the data dependencies.}
            }
            \label{fig:numerical}
        \end{figure}

        {
        To check whether the approximations used in the previous section regarding the negligible diffraction in the nonlinear crystal are valid, we support them with numerical simulations using COMSOL Multiphysics \cite{comsol}, where we perform nonlinear full-wave simulations in the undepleted pump approximation regime, similar to the numerical approach used in Refs.~\cite{fedotova2020,Yang2024}.
        Due to the large geometric size of the assumed nonlinear crystal in our analysis (compared to the wavelength of light), full-wave 3D calculations are computationally heavy, as they require a large number of mesh elements. Instead, a 2D system is considered for full-wave simulations as an illustrative example. The calculations are provided for a finite-sized nonlinear crystal with the transversal size $H$ equal to the total beam diameter $d_{\text{tot}}$. To allow a clear interpretation of the physics of diffraction and SHG, we model the nonlinear crystal as an isotropic medium with refractive index of $1.8$ at both the fundamental- and second-harmonic wavelengths of \(\lambda_\mathrm{FH}\)=1064 nm and \(\lambda_\mathrm{SH}\)=532 nm, respectively (as an approximate representation of the refractive indices of a KTP crystal along its different crystal axis in the visible and near-infrared wavelength ranges \cite{Kato:02}). To implement the free-space condition, the computational domain is surrounded by the perfectly matched layer (PML) as shown in Fig.~\ref{fig:numerical}A. The incident beam propagating along the $z$-axis (not to be mistaken with the crystal $z$-axis) is defined as a sum of $N$ spatially-shifted Gaussian functions with a random amplitude and an equal waist radius $w_0=H/2N$, where each Gaussian function represents a spatial feature/pixel:
        \begin{equation}
        E_\omega(x, z=0) = \sum\limits_{n=1}^{N} A_n \exp\left[ -\frac{(x - \Delta x_n)^2}{w_0^2} \right] \exp\left[-\frac{x^2}{\sigma^2 d_{\text{tot}}^2} \right], \end{equation}
        where $A_n$ are randomly selected amplitudes and $ \Delta x_n = -\frac{H}{2} + w_0 + 2(n-1) w_0$ is the shift of each Gaussian beam's center, such that the values of $x$ coordinate span from $-H/2$ to $H/2$. The second exponential factor, with $\sigma=0.2$, is added to smoothen the envelope of the distribution in such a way that the amplitudes near the ends of the crystal transversal interface are minimized while the maximum is located near $x = 0$. An example of the resulting fields from the full-wave simulations is shown in Fig.~\ref{fig:numerical}B.
        \\For our simulations, we fix the number of spatial features to $N=56$. Then, for each simulation, we choose a crystal transverse size $H$ and a $b$ value. Choice of $H$ determines $w_0$ ($H=d_{\mathrm{tot}} = N\Delta x=2Nw_0$), which consequently determines $z_R$ through eq.~\ref{eq:zR}. Then the choice of $b$ determines the crystal length $L$ through eq. \ref{eq:approximationQuality}, which eventually affects the quality of the negligible-diffraction approximation.} 
        
        {The aim is to rigorously determine the extent of the validity of the negligible-diffraction approximation, where the crystal performs the transformation $E_{2\omega}(x, z = L) \approx C E_{\omega}^2(x, z = 0)$. The proportionality constant $C$ is defined as
        \begin{equation}
        	C = \frac{\sum {E_{\omega}^*}^2 E_{2\omega}}{\sum|E_{\omega}^2|^2}.
        \end{equation}
        The similarity of the distributions is then quantified by the proportionality error 
        \begin{equation}
        	\xi = \frac{\sqrt{\sum\limits |E_{2\omega} - C E_{\omega}^2|^2}}{\sqrt{\sum\limits |E_{2\omega}|^2}},
        \end{equation}
        such that the summation for both $C$ and $\xi$ is performed over the $x$ coordinate.}
        
        {The proportionality error is calculated for three values of the transverse crystal size $H$, as a function of the $b$ coefficient, with the result shown in Fig.~\ref{fig:numerical}C. For generality, for each value of $H$, a different incident field distribution is used. It can be seen that the proportionality error is below $0.1\%$ for $b=1$, and as expected, decreases further with increasing the b coefficient, as the diffraction effects subside. It should be noted that the full-wave simulations also show some dependence of the overall error level on the total beam diameter. To further investigate this dependency, the calculations are provided for the case when the shape of the incident field is fixed, while all geometric parameters of the system are scaled. The reference system (with a scaling coefficient of 1) is considered to be a system with $H_\mathrm{ref} = 0.1$ mm, and simulations are performed for $H=\mathrm{scaling\: coefficient}\times H_\mathrm{ref}$. The crystal length is then determined as before from eq. \ref{eq:approximationQuality}. The calculated proportionality error is then calculated as a function of the scaling factor, for three different values of $b$, with the results shown in Fig.~\ref{fig:numerical}D. It can be see here that as expected, the $b$ coefficient has a strong role in setting the error level. Yet, the error also shows a systematic dependence on the scaling factor, where larger scaling factors, and hence a longer system, seems to exhibit a larger error based on the full-wave simulations. This is likely a numerical effect, as such full-wave simulations experience discretization-related numerical errors that accumulate over propagation distance, hence making simulated results for longer propagation distances deviate more from an exact solution \cite{HENNEKING202185}. Nonetheless, regardless of the origin of this dependency on the scaling factor, the full-wave simulations show that increasing $b$ can reduce the error, and hence improve the approximation, as was predicted analytically in the previous section, and the fact that even at a value of $b=1$, the error based on the negligible-diffraction approximation is below $0.1\%$.}

        
    \subsection{Aperture losses and focusing efficiency}
        We next analyze how the linear part of the DNN, made from consecutive phase modulation layers, affects the transmitted power and how efficiently that power can be concentrated at the detector.
        The phase modulation layer operates only on the lower spatial frequencies of the Fourier-transformed input, blocking the higher spatial frequencies, which are not covered by the phase modulation layer. This is reasonable, as the Fourier transformed image is very sparse at higher spatial frequencies. It is worth noting, that the transverse size of the phase modulation layers is an optimization parameter in our simulations, thus it will optimize to the area where the Fourier-transformed input-image information is spatially dense.
        As a result, a portion of the total power never participates in the nonlinear process nor contributes at the detection plane. 
        For the MNIST Digit dataset in the single layer DNNs as introduced in Fig.~\ref{fig:Setups-SingleLayer}, the transmitted power corresponds to  77.3\%--88.8\%  of the total power for the optimized phase modulation layers. 
        To quantify the focusing efficiency, we introduce a metric \(\epsilon =  I_c/I_{tot}\) based on the ratio of the intensity inside the detection patch corresponding to the correct class \(I_c\) with respect to the total optical intensity at the detection plane \(I_{tot}\). In our simulations, the focusing efficiency ranged from 1\%-26\% depending on SHG positioning and dataset in the single layer DNN. 
    \subsection{Exemplary output power and SNR estimation}
\begin{figure}[hbtp]
            \centering
            \includegraphics[width=1\linewidth]{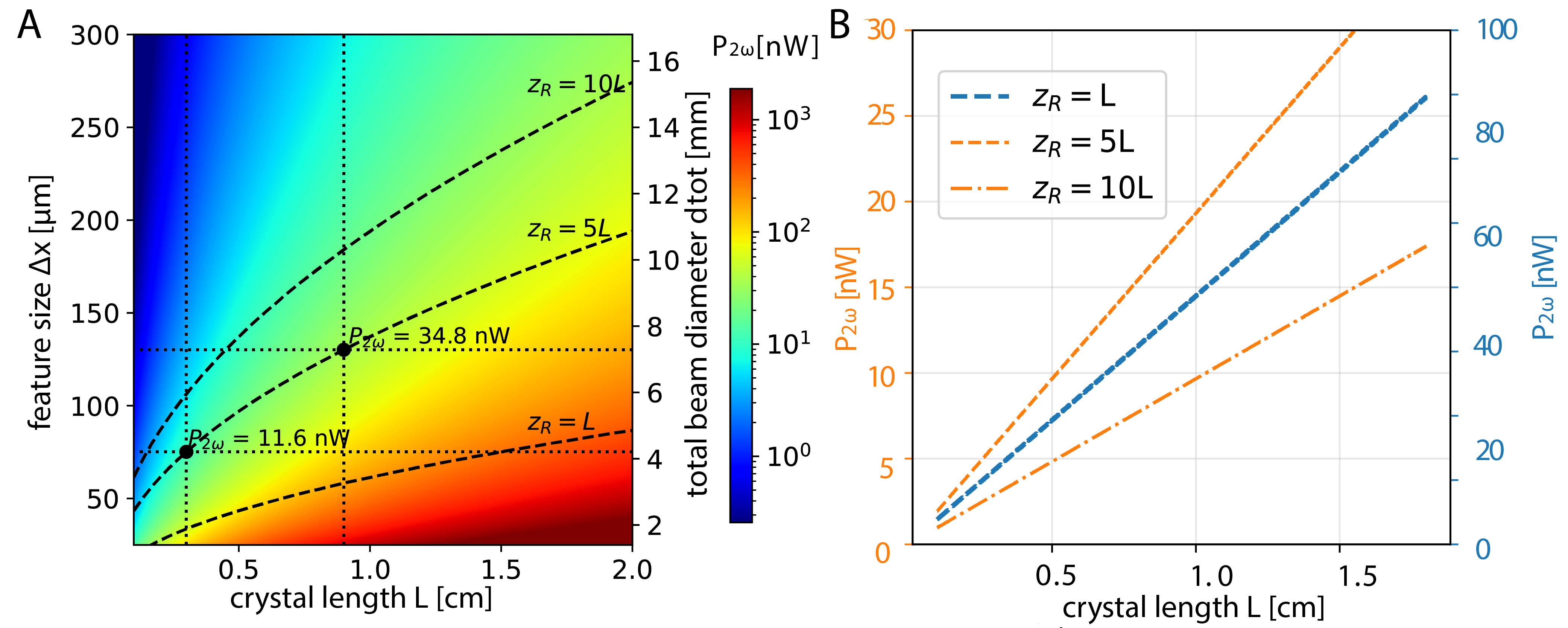}
            \caption{
            Expected total second-harmonic output power \(P_{2\omega}\) for an total total input fundamental harmonic power of \(P_{\omega}=1\)W as a function of crystal length \(L\) and feature size \(\Delta x\) (left axis) or total beam diameter \(d_{\mathrm{tot}} = \mathrm{N}\Delta x\)  (right axis) for \(\mathrm{N}^2=56\times56\) features. 
            A KTP-crystal, phase-matched for SHG and fundamental-harmonic wavelength of \(\lambda=1064\) nm , is assumed.
            The orange and blue dashed lines represent different levels to the approximation that there is no transverse mixing within the crystal, $L = z_R$, $5L = z_R$ and  $10L =  z_R$. 
            The black dotted lines mark a sample points for \(z_R = 5L\) at (\(L = 3~\mathrm{mm}, \Delta x = 143~\mathrm{\mu m}\)) and (\(L = 9~\mathrm{mm}, \Delta x = 247~\mathrm{\mu m}\)). }
            \label{fig:POWER}
        \end{figure}

        Based on our derivations above, we can now give a rough estimate of the output power fora given input power. Fig.~\ref{fig:POWER}(A) shows the estimated total output power for an input fundamental harmonic power of \(P_{\omega} = 1\)W into the crystal and a given crystal length \(L\) and feature size \(\Delta x\). The total beam diameter is defined as \(d_{\mathrm{tot}} = \mathrm{N}\Delta x\).  For simplicity we assume that the number of features in the total beam diameter corresponds to the number of pixels N in our phase modulation layer.
        Note that here, corresponding to our simulations, we assumed \(N=56\). This means that \(P_{\mathrm{\omega}} = N^2p_\omega\) and \(P_{\mathrm{2\omega}} = N^2p_{2\omega}\). The dashed lines in the plot correspond to different values of \(b\) in eq.~\ref{eq:approximationQuality}. Hence, choosing a larger
        \(b\) strengthens the condition of negligible transverse mode mixing within the crystal, meaning that the underlying approximation is better satisfied and diffraction effects are weaker.
        Fig.~\ref{fig:POWER}(B) shows the output power along these lines over the crystal length. 
        We can observe that the total output power \(P_{2\omega}\) increases linearly with crystal length, additionally it decreases with increasing approximation quality (values of b). That means there is an intrinsic trade-off.
        To find the output power for a given experimental setup, we can choose crystal length \(L\) and approximation quality measure \(b\), which also corresponds to a required feature size. In order to satisfy this feature size requirement, the crystal input can be magnified or de-magnified by an imaging scheme as illustrated in Fig.~\ref{fig:power_schematic}.
        Practical limitations on the crystal size must also be taken into account, both length and transverse size. The value for b should be chosen such that the total beam diameter is not larger than crystal transverse area size.
        Now, let us assume the single layer DNN with SHG at position 3 as introduced in Fig.~\ref{fig:Setups-SingleLayer}A.  
        
        For the crystal we assume a  
        potassium titanyl phosphate (KTP) crystal   \(n_{2\omega} =  n_{\omega} \approx 1.8\) 
        and an effective nonlinear coefficient $d_{\rm eff} = 3.5$~pm/V ~\cite{alford2001wavelength}.
        We select two sample points, for \(z_R = 5L\) at 
        \(L = 3~\mathrm{mm}\), resulting in \(\Delta x = 142.6~\mu\mathrm{m}\) and 
        \(P_{2\omega} = 11.6\mathrm{nW}\) and 
        \(L = 9~\mathrm{mm}\), resulting in 
        \( \Delta x = 246.9~\mu\mathrm{ m}\) and \(P_{2\omega} = 34.8 \mathrm{nW}\). This gives us the estimate of the conversion efficiency  of the SHG-process. As discussed above additionally there are aperture-related losses, for this specific DNN we have 88\% transmission through the linear part of the DNN and expect a focusing efficiency of \(\epsilon=\)19\%. If we are assuming \(P_{in}=1\)W is the input FH power emitted from the digits.
        Then, the detected output power is \(P_{\mathrm{out}}(L,\Delta x) = (0.88)^2P_{2\omega}(L,\Delta x) 0.19 \).  
        
        For \(L = 3~\mathrm{mm}\) and \(\Delta x = 143~\mu\mathrm{m}\), \(P_{2\omega}
        = 11.6~\mathrm{nW}\), yielding \(P_{\mathrm{out}} =  1.7~\mathrm{nW}\), which is the SH power that can be detected in a corresponding detection patch.  
        Similarly, for \(L = 9~\mathrm{mm}\) and \(\Delta x = 247~\mu\mathrm{m}\), \(P_{out} = 5.1~\mathrm{nW}\).

        \noindent{The resulting output powers are readily detectable by modern low-noise sCMOS camera systems. A detailed quantitative analysis of the SNR is provided in Section 8 of the Supplementary Materials. There, we show that for realistic operating conditions, we achieve a shot-noise-limited regime, yielding a signal-to-noise ratio of approximately 42 dB}~\cite{hamamatsu_C15440_20UP,Nakamura2006Basics}.
        \noindent{Note, that we chose an input power of \(P=1\,\mathrm{W}\) as a convenient reference point, such that the expected output power for any experimental implementation can be obtained by straightforward scaling according to
        \(P_{2\omega} = \eta P_{\omega}^2\).}
    
        {We point out that we performed the above power estimation assuming a continuous wave fundamental harmonic laser. In practice, one can significantly enhance the efficiency of the SHG process by using a pulsed laser excitation that delivers a higher peak fundamental harmonic power given the same amount of average power. As a rule of thumb, assuming a pulsed laser is chosen that its spectral bandwidth fits within the phase-matching bandwidth of the nonlinear crystal, and also assuming undepleted SHG without other non-ideal effects like saturation or crystal damage, the average generated SH power scales approximately proportional to the inverse of the duty cycle of the pulsed laser (duty cycle being equal to pulse temporal length multiplied by the repetition rate of the laser)~\cite{zipfel2003}. Another approach could be to use self-imaging nonlinear cavities, which benefit from resonances to enhance the intracavity intensity of the field, while preserving the spatial shape of the input beam profile~\cite{chalopin2010}. }
        
\section{Discussion and conclusion}
    In conclusion, our numerical simulations demonstrate that incorporating a {SHG activation} into a DNN, operating in the undepleted pump regime, can enhance classification accuracy and class contrast for single- and multi-layer DNNs and also for various classification problems.
    We observed a crucial dependence of the achievable classification accuracy on the position of the {SHG activation} in the DNN.  Depending on the {crystal} position, we showed that including SHG leads to increased or decreased classification performance. 
    For the MNIST digit classification in a single layer DNN, we observed a classification accuracy in the range of 89.6\% to 95.2\% depending on the SHG placement, compared to the linear case where we got a classification accuracy of 91.3\%.
    For the fashion digit classification in a 4-layer DNN, we observed a classification accuracy in the range of 80.9\% to 85.7\% depending on the SHG placement, compared to the linear case where we got a classification accuracy of 84.2\%.
    We found a consistent improvement when the {SHG activation} is placed with some propagation distance after the last phase modulation layer.   
    Additionally, we observed that adding SHG into the DNN can bypass the accuracy-contrast trade-off ~\cite{fan2025joint}.
    In the single layer DNN, we observed a class contrast enhancement from 32.1\% to 56.0\% for the MNIST-digit, from 38.1\% to 55.4\% for the MNIST-fashion and from 5.9\% to 7.6\% for the EMNIST dataset.
    
    As our results demonstrate, SHG in the undepleted regime can act as a useful nonlinear activation layer for DNNs. Such a polynomial nonlinearity has the special advantage that its quadratic functionality does not depend on the absolute level of power. That means it can work in principle for arbitrary small input optical powers if a sufficiently sensitive photodetector is used at the output plane. This could be a major advantage for the implementation of power-efficient all-optical nonlinear DNNs.  
    At the same time, it should be pointed out that a single undepleted {SHG activation}, as investigated in our work, does not result in the universal approximation capability. Nevertheless, SHG aligns well with a broader mathematical framework for function approximation. Polynomial networks have demonstrated strong capabilities in function fitting and feature extraction \cite{zhou2019, kileel2019}. According to the Weierstrass theorem \cite{stone1948}, any continuous function can be approximated by polynomials, and many functions can be effectively represented using only the first few polynomial orders. This makes SHG a promising candidate for future developments in optical computing, if one could efficiently cascade multiple layers incorporating SHG to increase the network depth. This is generally challenging in practice, due to efficiency constraints with repeated frequency doubling. However, the emergence of nanophotonic solutions, such as nonlinear metasurfaces \cite{zheng2023,fedotova2024}, creates promises for the realization of thin, yet highly efficient nonlinear elements for DNNs. Of specific interest are works exploiting metasurfaces for nonlinear imaging \cite{del2021} and for generating multiple higher harmonics \cite{zograf2022}.
    
    Finally, we emphasize that realizing a pointwise SHG response of the form \(E_{2\omega}(x,y) = E_\omega(x,y)^2\) in the nonlinear
    crystal, requires that diffraction inside the nonlinear medium is negligible. We show that this condition results in an intrinsic trade-off between crystal length and beam diameter: the crystal must be shorter to avoid transverse mixing, yet long enough to yield more second-harmonic output. Within these constraints, we estimate the output power for realistic parameters. The result indicates that, for an input power of \(1\)W we can achieve signal strengths that are detectable with standard photo detectors, even after accounting for losses associated with the phase modulation layer and the fraction of light contributing to the correct class. Overall, our results suggests a path for including SHG into DNNs, as a feasible nonlinear mechanism leading to improved classification contrast and accuracy.

    \section*{Funding.}
    \addcontentsline{toc}{section}{Funding}
    Carl Zeiss Foundation (P2022-04-018, 0563-2.8/738/2); Deutsche Forschungsgemeinschaft (437527638, 398816777, 505897284).
    \section*{Acknowledgments.}
    \addcontentsline{toc}{section}{Acknowledgments}
        This work was made possible by funding from the Nexus program of the Carl-Zeiss-Stiftung
        (project MetaNN, project ID P2022-04-018). The authors also acknowledge funding from the Carl-Zeiss-Stiftung under the project ID 0563-2.8/738/2 (Breakthrough project "A Virtual Werkstatt for Digitization in the Sciences"), and the German Research Foundation (DFG) under the project numbers 437527638 (IRTG 2675 Meta-Active), 398816777 (CRC/SFB 1375 NOA - Nonlinear Optics down to Atomic scales), and 505897284 (MEGAPHONE project).
    \section*{Disclosures.}
    \addcontentsline{toc}{section}{Disclosures}
        The authors declare no conflict of interest.
    \section*{Data availability.}
    \addcontentsline{toc}{section}{Data availability}
        Data underlying the results presented in this article may be obtained from the authors upon reasonable request.
    \section*{Supplementary Materials.}
    \addcontentsline{toc}{section}{Supplementary Materials}
        See Supplement 1 for supporting content.

\bibliographystyle{opticajnl}
\bibliography{references}
\section*{\LARGE \textbf{Supplementary Materials}}
\addcontentsline{toc}{section}{Supplementary Materials}
\appendix

    \section{Implementation of Free-Space Propagation}

    In our system, light propagates between two parallel planes centered along the optical axis, separated by a distance \( z \), as illustrated in Fig.~\ref{fig:overview}. For numerical implementation, we adopt the method introduced in ~\cite{shen2006}, which solves the Rayleigh-Sommerfield integral:
    \begin{equation}\label{eq:supp1}
    E(x,y,z) = E_0(x', y') * h(x', y', z) 
    \end{equation}
    The integral computes the electric field \( E(x,y,z) \) at plane \( z \) using the free-space response function that connects points between the aperture and observation planes:
    \begin{equation}\label{eq:supp2}
        h(x', y', z) = -\frac{1}{2\pi} \frac{z}{R^2} \left( jk - \frac{1}{R} \right) e^{jkR} 
    \end{equation}
    \begin{equation}
        R = \sqrt{(x' - x)^2 + (y' - y)^2 + z^2} 
    \end{equation}
    
    Here, \( k \) is the wavenumber in free space. Physically, this represents each point on the aperture plane acting as a secondary source of a spherical wave. The integral in Eq.~(\ref{eq:supp1}) is evaluated via the convolution theorem, where  the convolution of the input field with the free-space response function is computed in the Fourier domain. Specifically, this involves multiplying the numerically Fourier-transformed input field by the Fourier-transformed response function.
    Note that the diffractive model does not rely on small-angle approximations (e.g., Fresnel diffraction) or far-field conditions (e.g., Fraunhofer diffraction), providing a general framework for wave propagation.
    Within the context of a diffractive neural network, this equation can be reformulated as
    \begin{equation}
        w_{mi}^l(r_{mi}^l) = \frac{1}{2\pi} \frac{z_{l-1} - z_l}{(r_{mi}^l)^2} \left( jk - \frac{1}{r_{mi}^l} \right) e^{jkr_{mi}^l}, 
    \end{equation}
    \begin{equation}
        r_{mi}^l = \sqrt{(x_m - x_i)^2 + (y_m - y_i)^2 + (z_{l-1} - z_l)^2}, 
    \end{equation}
    where \( w_{mi}^l \) corresponds to the weight connecting neuron \( m \) in layer \( l-1 \) to neuron \( i \) in layer \( l \). The output field at layer \( l \) and neuron \( i \) is modeled as
    \begin{equation}
        u_i^l(x_i, y_i, z_l) = t_i^l(x_i, y_i, z_l) \sum_m w_{mi}^l(r_{mi}^l) \, u_m^{l-1}(x_m, y_m, z_{l-1}) \, \Delta x_m \Delta y_m. 
    \end{equation}
    Here, \( u_i^l(x_i, y_i, z_l) \) denotes the output of neuron \( i \) in layer \( l \), located at the spatial coordinates \( (x_i, y_i, z_l) \). The sum \( \sum_m \) represents the contribution of all neurons in the previous layer, where \( \Delta x_m \Delta y_m \) corresponds to the area of a neuron.
    The phase modulation at layer \( l \) is denoted by the complex transmission coefficient:
    \begin{equation}\label{eq:supp_transmission_coefficient}
        t_i^l(x_i, y_i, z_l) = e^{j\Phi_i^l(x_i, y_i, z_l)} 
    \end{equation}
    To ensure numerical stability, the computational grid resolution was set to half a wavelength. 

\section{Training Parameters} 

    The training parameters and the loss function in our model were defined analogously to the framework proposed by ref.~\cite{mengu2020}. In the DNN framework, each neuron is characterized by the complex transmission coefficient described in eq.~\ref{eq:supp_transmission_coefficient}. Phase modulation is represented by the physical parameter \( \Phi_i^l \). During training, it is represented by the latent trainable variable \( \beta_i^l \):
    
    \begin{equation}
        \Phi_i^l = 2\pi \beta_i^l
    \end{equation}
    
    Note that while the phase is in principle unbounded in this formulation, the inherent periodicity of \( e^{j\Phi} \) enables the backpropagation to converge effectively. The trainable parameters are optimized using the backpropagation method with the Adam optimizer and a learning rate of \( 10^{-3} \)  ~\cite{kingma2014adam} .
    Note that in our network, we considered layers only modulating the phase, assuming complete transmissivity and no amplitude modulation. The input into our network was encoded purely in amplitude.

\section{Loss} 

    For the loss \( L \), we applied the sparse categorical cross-entropy: 
    \begin{equation}\label{eq:SCA}
        L = -\sum_{c=1}^C y_c \log(\hat{y}_c)
    \end{equation}
    where \( C \) is the number of classes, \( y_c \) is the true label (1 for the correct class and 0 otherwise), and \( \hat{y}_c \) is the predicted probability for class \( c \). That means that the final diffractive design relies solely on the maximum optical signal detected at the output plane for class assignment.
    In our implementation, the intensity measured at the patches for each class (areas \( p_c \) on the detection plane) is integrated to form a measurement vector \( \mathbf{M} \) of length \( C \):
    \begin{equation}
        \begin{array}{rcl}
            \mathbf{M} &=& \left( \int_{p_1} I(x, y), \int_{p_2} I(x, y), \ldots, \int_{p_C} I(x, y) \right)^T \\[0.5em]
                       &=& (m_1, m_2, \ldots, m_C)^T
        \end{array}
    \end{equation}
    We normalized the measurement vector by its maximum to make the problem independent of patch size or sampling points:
    \begin{equation}
        \hat{\mathbf{M}} = \frac{\mathbf{M}}{\max(\mathbf{M})}
    \end{equation}
    Since cross-entropy operates on probability measures (values in the interval (0,1)), a softmax function is applied to the measurement vector to produce normalized probabilities:
    \begin{equation}
        \hat{y}_c(\hat{\mathbf{M}}) = \frac{e^{m_c}}{\sum_{j=1}^C e^{m_j}}
    \end{equation}

\section{Numerical Parameters and Their Physical Meaning} 
    In this section, we provide a comprehensive overview of the key parameters used in our optical simulation setup and how they relate to real physical quantities. Our implementation employs two numerical Fourier transforms and is crucial to ensure that the results are physically meaningful and realizable in an actual optical system. 
    \begin{figure}[htbp]
        \centering
        \includegraphics[width=0.9\linewidth]{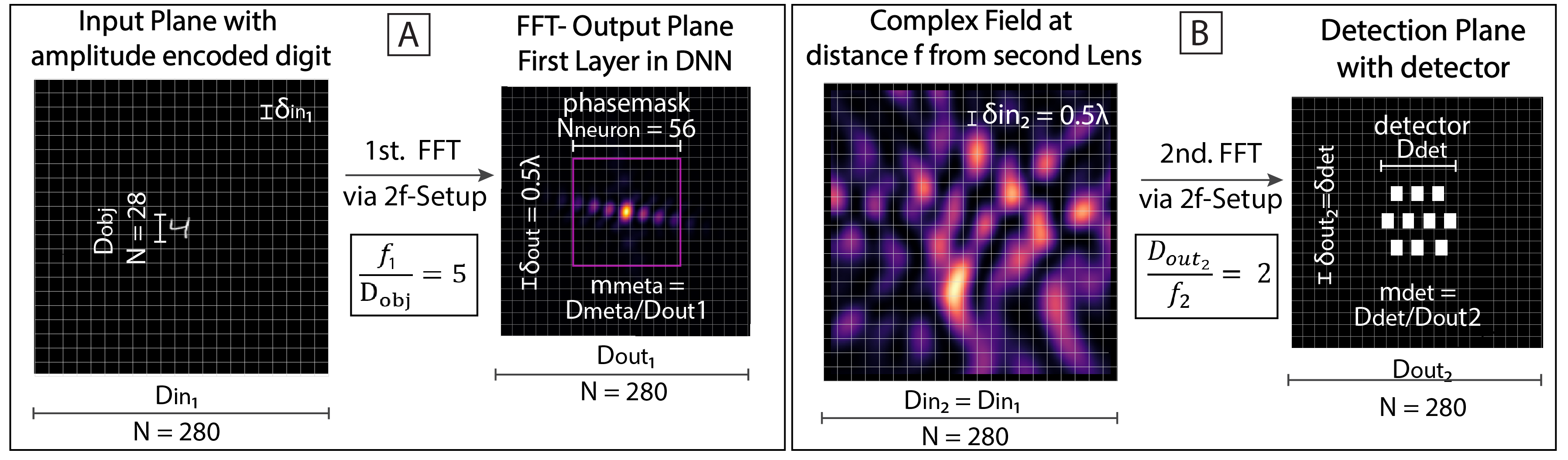}
            \caption{Overview of numerical parameters used in the simulations as well as their corresponding physical parameters. (A) Parameters governing the Fourier transform between the object and the first phase modulation layer, and (B) those defining the transformation from the final plane (either the last phase modulation layer or the propagated field to the detector.}
        \label{fig:overview}
    \end{figure}
    \paragraph{Overview of the Numerical Fourier Transform}
        \noindent Fig.~\ref{fig:overview} provides a schematic overview, where Fig.~\ref{fig:overview}(A) shows parameters related to the Fourier transform between the object and the first phase modulation layer, and Fig.~\ref{fig:overview}(B) shows parameters for the Fourier transform from the final plane (either the last phase modulation layer or a propagated field) to the detector. The intermediate neural network inherits the spatial and spectral properties defined by the first transform.
        Both Fourier transforms in our system are numerically equivalent. The number of sampling points \(N\) connects the simulation to several key physical parameters, including the lateral size of the output plane \(D_{\text{out}}\), the input sampling interval \(\delta_{\text{in}} = \frac{D_{\text{in}}}{N}\), the lateral size of the input plane \(D_{\text{in}}\), and the focal length of the Fourier-transforming lens \(f\).
        These relationships are governed by Fourier optics:
        \begin{equation}
            D_{\text{out}} = \frac{\lambda f}{\delta_{\text{in}}} = \frac{N \lambda f}{D_{\text{in}}}
        \end{equation}
        or equivalently,
        \begin{equation}\label{eq:supp14}
            \delta_{\text{out}} = \frac{\lambda f}{N \delta_{\text{in}}}
        \end{equation}
        To simplify computations and generalize applicability, we set the wavelength \( \lambda = 1 \). This means that all spatial dimensions are expressed in units of wavelength, allowing the trained model to work with any monochromatic light source, if parameters are scaled accordingly.
    
    \paragraph{Parameter Choices for the First Fourier Transform (Fig.~\ref{fig:overview}(A)}
        \noindent Our training input data consistently has a resolution of \( 28 \times 28 \) pixels. We pad this to \( 280 \times 280 \) to ensure a smooth and well-resolved Fourier transform. Hence, \( N = 280 \). The output sampling after the first transform is fixed to \( \delta_{\text{out}_1} = \delta_{\text{meta}} = \lambda / 2 \), ensuring accurate sampling for further propagation. Plugging this into eq.~\ref{eq:supp14} leads us to:
        \begin{equation}
            \frac{f}{N \delta_{\text{in}_1}} = \left( \frac{f_1}{D_{\text{obj}}} \right) \left( \frac{N_{\text{obj}}}{N} \right) = 0.5,
        \end{equation}
        \begin{equation}
            \frac{f_1}{D_{\text{obj}}} = 5.
        \end{equation}
        Thus, any combination of focal length \( f_1 \) and object size \( D_{\text{obj}} \) that satisfies \( f_1 / D_{\text{obj}} = 5 \) will produce the simulated output.

    \paragraph{Parameter Choices for the Second Fourier Transform (Fig.~\ref{fig:overview}(B)))}
        \noindent From the previous layer, we inherit \( \delta_{\text{out}_1} = \delta_{\text{in}_2} = \lambda / 2 \) and \( N = 280 \). Plugging this into eq.~\ref{eq:supp14} leads us to the following:
        \begin{equation}
            \quad \frac{\delta_{\text{out}_2}}{f_2} = \frac{\lambda}{N \delta_{\text{in}_2}} = \frac{1}{140},
        \end{equation}
        or
        \begin{equation}
            \frac{D_{\text{out}_2}}{f_2} = \frac{\lambda}{\delta_{\text{in}_2}} = 2. \label{eq:Ddet}
        \end{equation}
        Thus, any lens with focal length \( f_2 \) will produce an output plane of size \( D_{\text{out}_2} = 2 f_2 \). As we do not necessarily cover the whole numerical Fourier transform with our detector patches and to ensure sufficient sampling of the field at the output layer, it makes sense to relate this equation to the output pixel size as displayed above. Note that in an experimental setup there is always the option to add more lenses to magnify or demagnify the input or output planes to the desired sizes.

    \paragraph*{Other Parameters}
        \noindent For the phase modulation layer, we set the number of neurons to \( 56 \times 56 \). The neuron size \( \delta_{\text{neuron}} = \Delta x = \Delta y \) was not fixed, but can in general be as small as the sampling size or span multiple sampling points. As described in the next section, the phase modulation layer size, and thus neuron size, is a parameter for which we optimized. The fraction of the Fourier transformed image that is covered by the phase modulation layer has great impact on the final performance. We describe this optimization parameter as
        \begin{equation}
        m_{\text{meta}} = \frac{D_{\text{meta}}}{D_{\text{out}_1}} = \frac{D_{\text{meta}}}{140\lambda}
        \end{equation}
        with \( \delta_{\text{neuron}} = D_{\text{meta}} / 56 \). We block any light that goes beyond the area of the phase modulation layer, meaning the phase modulation layer acts as an aperture. Note that in the multilayer DNN, the parameters of the first phase modulation layer, i.e., neuron size and number, are inherited by the other layers.
        Additionally, we also optimize the size of the detector, described as
        \begin{equation}
            m_{\text{det}} = \frac{D_{\text{det}}}{D_{\text{out}_2}}\label{eq:mout}
        \end{equation}
        Overall, \( m_{\text{meta}} \) and \( m_{\text{det}} \) are two hyperparameters for our DNN. According to~\cite{Zheng:22}, setting the Fresnel number (which connects the sampling size to the propagation length) within \( F \in (2 \times 10^{-4}, 1 \times 10^{-2}) \) maximizes the expressiveness of the DNN. We set the inter layer propagation distance to \( z = 30 \lambda \), which corresponds approximately to Fresnel numbers on the order of
        \begin{equation}
            F = \frac{\delta_{\text{neuron}}^2}{\lambda z} = \frac{\Delta x^2}{\lambda z} \approx 10^{-2}
        \end{equation}
        with \(\Delta x = \dfrac{D_{\text{out}_1} \cdot m_{\text{meta}}}{56} = \delta_{\text{out}_1} \cdot \dfrac{N}{56} \cdot m_{\text{meta}} = 0.5\lambda \cdot \dfrac{280}{56} \cdot m_{\text{meta}} = 2.5\, m_{\text{meta}}\, \lambda \)
        
    \section{Optimization Parameters and their Impact on Performance} 

         \begin{figure}[htbp]
            \centering
            \includegraphics[width=1\linewidth]{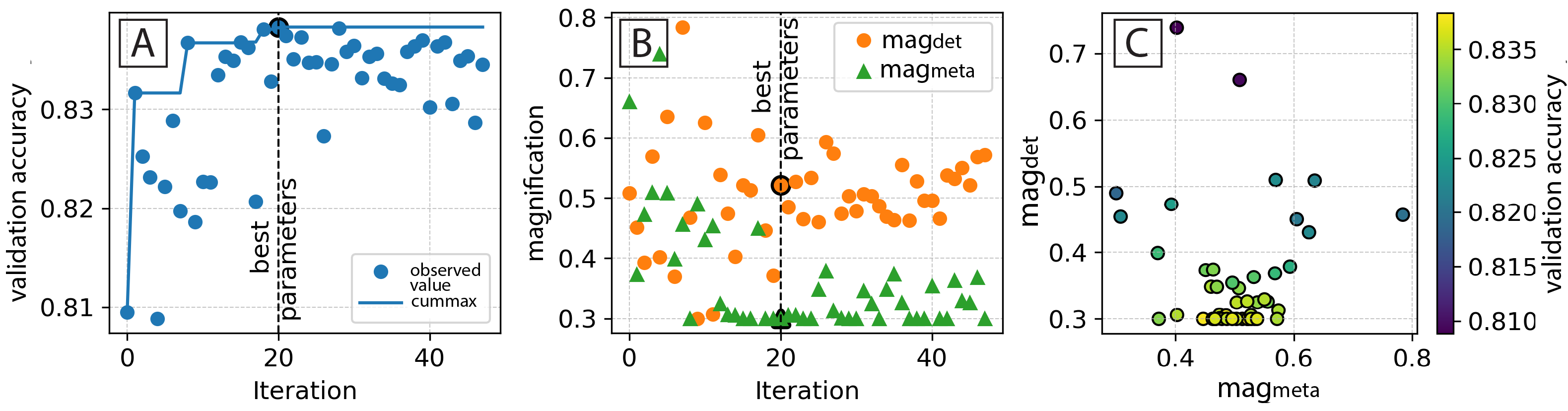}
            \caption{Exemplary convergence behavior of the optimization parameters for the multilayer DNN without SHG on the fashion MNIST dataset. (A) The blue points represent the validation accuracy at each iteration, the blue line shows the cumulative maximum (cummax), (B) shows the corresponding parameters at each iteration and (C) shows the sampled parameters with respect to the validation accuracy.}
            \label{fig:optimization_convergence}
        \end{figure}
        
        The performance of a Diffractive Neural Network (DNN) is highly dependent on its system parameters. There are always cases where one configuration may outperform another, depending on the chosen parameters. To ensure a fair comparison and allow each DNN configuration to perform at its best, we optimize the two hyper parameters $m_{\text{det}}$ and $m_{\text{meta}}$ separately for each individual DNN configuration.
        For optimization, we applied a Bayesian optimizer ~\cite{nogueira2014}, which models the objective function probabilistically using Gaussian processes to balance exploration and exploitation of the optimization space efficiently. In combination, we used sequential domain reduction ~\cite{stander2002}, which iteratively narrows the search space around promising regions to improve and speed up the convergence of Bayesian optimization.
        Specifically, the validation accuracy was maximized, which was evaluated after three epochs of training before testing the next set of hyper-parameters.
        As described above, in our DNN, we apply a Fourier transform to the input image at the first layer and likewise Fourier transform the output plane onto the detector. Here we optimize for two parameters $m_{\text{meta}}$ and $m_{\text{det}}$ as defined above.
        $m_{\text{meta}}$ is a key factor that influences the performance of the DNN when the size of the neurons is limited. Ideally, it would always be favorable to squeeze as many neurons into an area as possible, up to the diffraction limit. However, for the tested datasets, most of the information in the Fourier-transformed image is concentrated in the lower spatial frequencies, located near the center, while the higher spatial frequencies approach zero at the edges. Consequently, contributions from these outer regions become increasingly weaker. As a result, the phase modulation layer naturally functions as a low-pass filter in addition to its modulation role. Therefore, it is more effective to position neurons (phase-modulation elements) in areas of high relevance. Note that the relevant spatial frequencies are strongly dependent on the dataset.
        Also note that the light outside of the phase modulation layer is blocked. This can be done in practice, for example, by metal-coating the area outside the phase modulation layer. In the 4-layer DNN, we maintained a constant size for the phase modulation layer.
        Fig.~\ref{fig:optimization_convergence} illustrates the convergence behavior of the parameters in an optimization run for a DNN configuration. In this example, we maximized the validation accuracy for the MNIST-fashion dataset in the 4-layer DNN over 50 iterations. The blue points show the achieved validation accuracy, the blue line corresponds to the cumulative maximum, and the points in the plot below correspond to the parameters sampled at each iteration. We observe that the optimizer successfully converges in validation accuracy. Once the optimal parameters were identified, training was resumed for 50 epochs using these parameters. The optimization parameters for each DNN configuration, as described in the manuscript, are summarized in Table ~\ref{tab:optimization_parameters}.

        \begin{table}[htbp]
        \centering
        \caption{Optimization Parameters}
            \begin{tabular}{|c|c|c|c|c|}
                \hline
                \textbf{Layers} & \textbf{Dataset} & \textbf{SHG Position} & \textbf{$m_{\text{meta}}$} & \textbf{$m_{\text{detector}}$} \\                  \hline
                1 & digits  & None & 0.42 & 0.40 \\                \hline
                1 & digits  & 1    & 0.31 & 0.49 \\                \hline
                1 & digits  & 2    & 0.30 & 0.61 \\                \hline
                1 & digits  & 3    & 0.33 & 0.30 \\                \hline
                1 & digits  & 4    & 0.40 & 0.40 \\                \hline
                1 & fashion & None & 0.60 & 0.30 \\                \hline
                1 & fashion & 3    & 0.32 & 0.30 \\                \hline
                1 & emnist  & None & 0.31 & 0.64 \\                \hline
                1 & emnist  & 3    & 0.30 & 0.74 \\                \hline
                4 & fashion & None & 0.47 & 0.30 \\                \hline
                4 & fashion & 1    & 0.41 & 0.041 \\                \hline
                4 & fashion & 2    & 0.41 & 0.30 \\                \hline
                4 & fashion & 3    & 0.69 & 0.30 \\                \hline
                4 & fashion & 4    & 0.64 & 0.30 \\                \hline
                4 & fashion & 5    & 0.62 & 0.47 \\                \hline
                4 & fashion & 6    & 0.56 & 0.35 \\                \hline
            \end{tabular}
            \label{tab:optimization_parameters}
        \end{table}

    \section{Training} 
        \begin{figure}[htbt]
            \centering
            \includegraphics[width=0.5\linewidth]{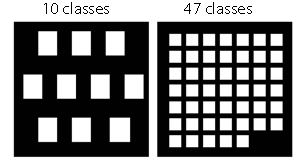}
            \caption{Target detector masks used for training of the DNN: (A) detector mask for the 10-class datasets (digit-MNIST and fashion-MNIST). (B) detector mask for the 47-class EMNIST Letters dataset. The masks define the spatial target patterns used during training to encode class labels into spatially localized intensity patterns.}
            \label{fig:detector_masks}
        \end{figure}
\begin{figure}[htbp]
            \centering
            \includegraphics[width=0.7\linewidth]{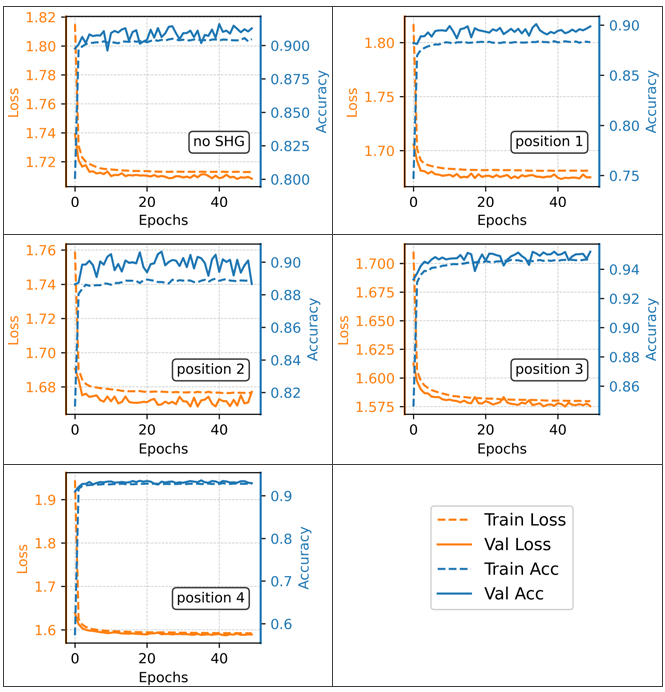}
            \caption{Training history of loss and accuracy for the single-layer DNN with 56×56 neurons per layer on the MNIST Digit dataset. The plots correspond to the single-layer setups 1–4 with SHG and the setup without SHG, as illustrated.}
            \label{fig:training_history_singleLayer}
        \end{figure}
\begin{figure}[htbp]
            \centering
            \includegraphics[width=0.7\linewidth]{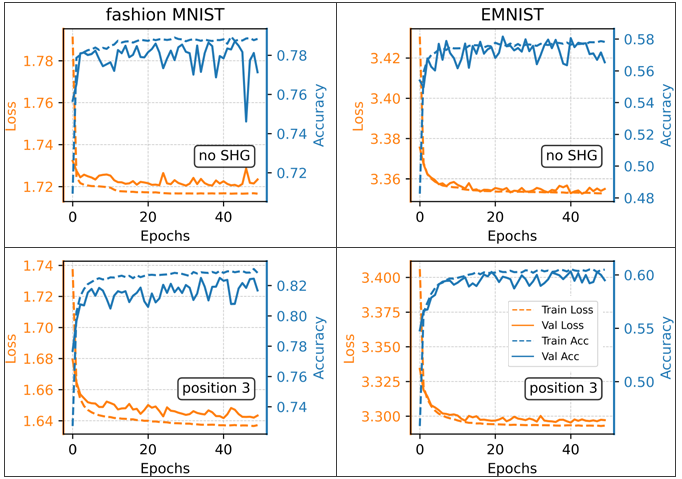}
            \caption{Training history of loss and accuracy for the single-layer DNN with 56\(\times\)56 neurons per layer on the MNIST Fashion Dataset (left) and EMNIST handwritten letters dataset (right) for a DNN without SHG (top row) and a DNN with SHG corresponding to SHG position 3 as introduced in Fig.~2 of the main text.}
            \label{fig:training_history_singleLayer_datasets}
        \end{figure}
\begin{figure}[htbp]
            \centering
            \includegraphics[width=0.7\linewidth]{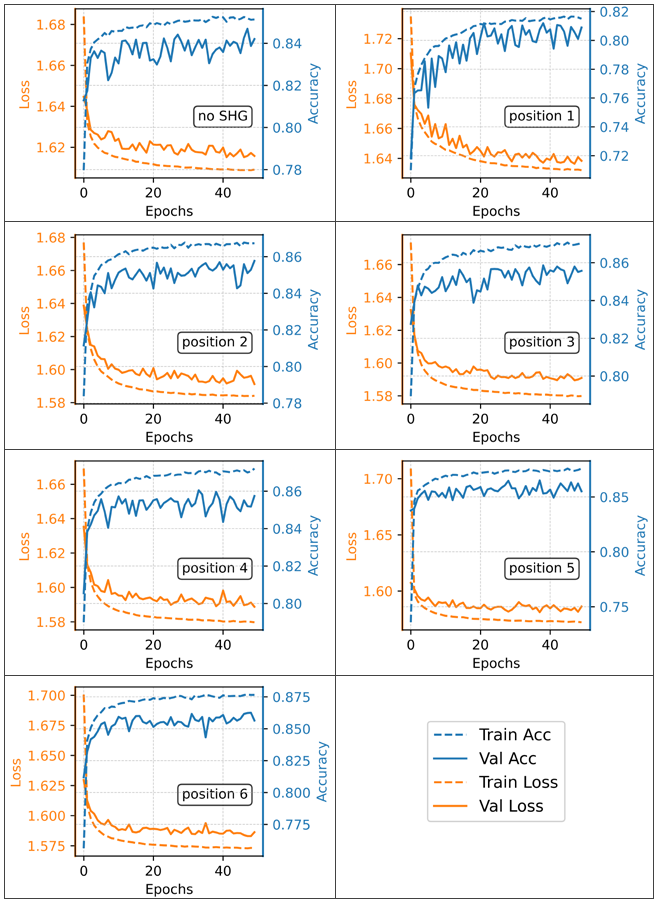}
            \caption{Training history of loss and accuracy for the 4-layer DNN with 56\(\times\)56 neurons per layer on the MNIST fashion dataset. The plots correspond to SHG position 1-6 of the 4-layer DNN  as introduced in Fig.~4 of the main text.}
            \label{fig:training_history_multiLayer}
        \end{figure}

        After parameter optimization for our analysis, we trained the DNN for both the single-layer and the 4-layer DNN each for 50 epochs. For both the MNIST digit  and MNIST fashion datasets, the training data were provided by the Keras package, consisting of 60,000 grayscale images of size $28 \times 28$ for training and 10,000 for validation, with 10 classes each ~\cite{deng2012mnist,xiao2017fashion}. The EMNIST handwritten digit data consists of 112,800 grayscale images for training and 18,800 for validation, with 47 classes and an image size of $28 \times 28$  ~\cite{cohen2017}. The phase modulation layers were optimized using the Adam optimizer in Keras, with the loss function defined as sparse categorical cross-entropy eq.~\ref{eq:SCA}. Training was conducted with a batch size of 8 and a learning rate of 0.001. In Fig.~\ref{fig:training_history_singleLayer} - Fig.~\ref{fig:training_history_multiLayer}, we provide the full training histories, including both training and validation loss and accuracy. In particular, as system complexity increases, whether through additional layers or more complex datasets, the convergence rate tends to slow.
        Fig.~\ref{fig:training_history_singleLayer} - Fig.~\ref{fig:training_history_multiLayer} display the loss and accuracy over training epochs (that is, one full cycle through the training images). For comparison purposes, we consistently used the final validation accuracy after 50 epochs. Note that validation accuracy naturally exhibits some fluctuations; in our case, these were typically within $\pm 0.5\%$ or less.
        Successful DNN training requires a target detector mask. Fig.~\ref{fig:detector_masks}(A) shows the mask used for the 10-class digit MNIST and fashion MNIST datasets, while Fig.~\ref{fig:detector_masks}(B) presents the mask used for the 47-class EMNIST handwritten letters dataset.

        
    \section{Focusing efficiency}\label{sec:focusingEfficiency}
\begin{figure}[htbp]
            \centering
            \includegraphics[width=0.9\textwidth]{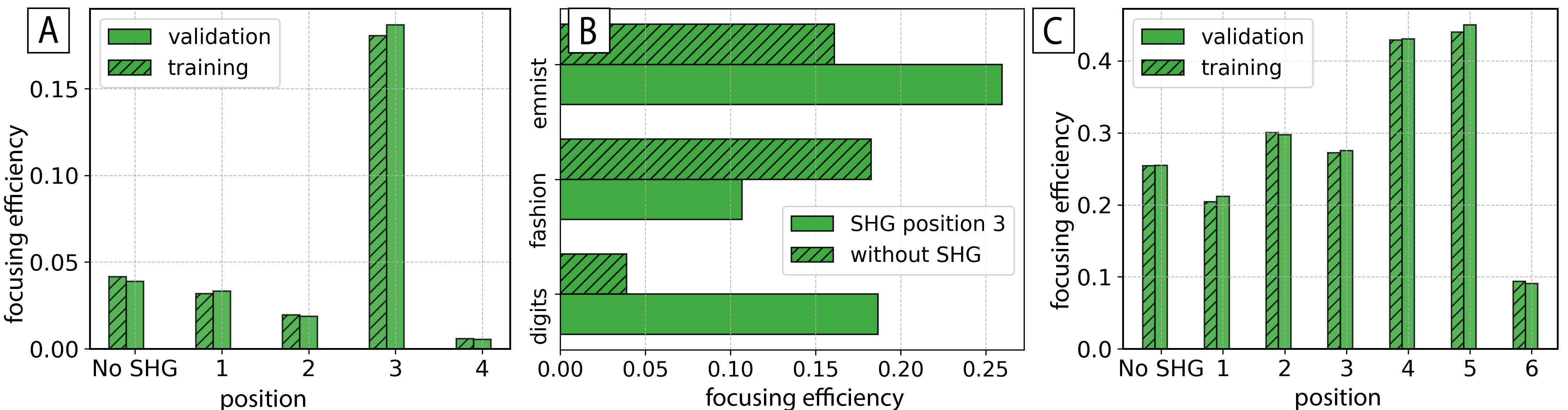}
            \caption{Focusing efficiency (averaged over all classes) for different DNN configurations:
             (A) Single-layer DNN with SHG at different positions, corresponding to Fig.~2 in the main manuscript (trained on the MNIST dataset). (B) Single-layer DNN across different datasets with SHG fixed at position 3. (C) Multi-layer DNN with SHG at different positions on trained on the Fashion-MNIST dataset, corresponding to Fig.~4 in the main manuscript.}
            \label{fig:focusingEfficiency}
        \end{figure}

        The focusing efficiency, defined as
    \begin{equation}
        \epsilon = \frac{I_c}{I_{\mathrm{tot}}},
    \end{equation}
    quantifies the fraction of total optical intensity that is directed into the correct detection region, where $I_c$ is the intensity inside the target patch and $I_{\mathrm{tot}}$ is the total intensity at the detection plane. Fig.~\ref{fig:focusingEfficiency} summarizes the focusing efficiency for several DNN configurations, extending the results presented in the main manuscript.
    
    \section{Detector Specifications and Signal-to-Noise Analysis}
    \label{sec:detector_snr}
    
    In this section we provide a brief discussion on detector specifications required for reliable operation of the proposed system.
    Modern cameras exhibit high quantum efficiency, and signal levels at the magnitudes encountered in our system should not pose a fundamental challenge.
    Here we estimate the signal-to-noise ratio (SNR) for an exemplary sCMOS camera (Hamamatsu C15440-20UP), which has a quantum efficiency of $\text{QE} \approx 92\%$ at $\lambda = 532\,\text{nm}$ and a pixel size of $D_\text{pixel} = 6.5\,\mu\text{m}$~\cite{hamamatsu_C15440_20UP}.
    The estimation proceeds in three steps: we first determine the power incident per detector pixel, from which we derive the number of detected electrons per frame, and finally compute the SNR.
    
    \paragraph{A) Power per Pixel.}
    A major factor governing the per-pixel signal is the spatial extent of the detector mask and its subdivision into patches.
    As discussed above, the diameter of the detector mask $D_\text{det}$ is determined by the focal length $f_2$ of the second lens and the optimization parameter $m_\text{out}$, which controls what fraction of the output plane the detector mask spans (see eq.~\ref{eq:Ddet} and eq.~\ref{eq:mout}):
    \begin{equation}
        D_\text{det} = m_\text{out} \cdot 2f_2.
    \end{equation}
    The detector mask can be arbitrarily chosen, with the only constraint that the number of patches matches the number of classes. 
    For a rough estimation we use the parameters of our system, in which each patch covers approximately $D_\text{patch} \approx 0.15 \cdot D_\text{det}$.
    With $m_\text{out} = 0.3$ this gives:
    \begin{equation}
        D_\text{patch} = m_\text{out} \times 0.15 \times 2f_2.
    \end{equation}
    Choosing an exemplary focal length of $f_2 = 2.5\,\text{cm}$ yields $D_\text{patch} \approx 2.25\,\text{mm}$.
    With a pixel size of $D_\text{pixel} = 6.5\,\mu\text{m}$, the power per pixel is:
    \begin{equation}
        P_\text{pixel} = \left(\frac{D_\text{pixel}}{D_\text{patch}}\right)^{2} \cdot P_\text{patch}
        = 4.27 \times 10^{-14}\,\text{W},
    \end{equation}
    given $P_\text{patch} = 5.12\,\text{nW}$.
    
    \paragraph{B) Signal-to-Noise Ratio.}
    The incident photon flux per pixel is:
    \begin{equation}
        \Phi = \frac{P_\text{pixel} \cdot \lambda}{hc} = 1.14 \times 10^{5}\,\text{photons\,s}^{-1},
    \end{equation}
    where $h$ is Planck's constant and $c$ the speed of light.
    The number of detected electrons is:
    \begin{equation}
        N = \text{QE} \times \Phi \times t \approx 1.47 \times 10^{4}\,\text{e}^{-},
    \end{equation}
    for $\text{QE} \approx 92\%$, $\lambda = 532\,\text{nm}$, and an integration time of $t = 0.14\,\text{s}$.
    This is just at the full well capacity of the camera ($\approx 1.5 \times 10^{4}\,\text{e}^{-}$), for weaker signals integration time could be easily increased.
    The photon detection follows Poisson statistics, thus the shot noise is $\sigma_\text{shot} = \sqrt{N} \approx 121\,\text{e}^{-}$, which exceeds the read noise ($\sigma_\text{read} = 0.7\,\text{e}^{-}$) and dark current ($\sigma_\text{dark} = 1.0\,\text{e}^{-}\,\text{pixel}^{-1}\,\text{s}^{-1}$) well enough, such that the SNR can be approximated by the shot-noise-limited expression \cite{Nakamura2006Basics,hamamatsu_db}:
    \begin{equation}
        \text{SNR} \approx 20 \cdot \log_{10}\!\left(\frac{N}{\sigma_\text{shot}}\right)
        = 20 \cdot \log_{10}\!\left(\sqrt{N}\right)
        \approx  42\,\text{dB}.
    \end{equation}
    This confirms that the optical power levels in the proposed system are well within the measurable range of a standard sCMOS camera for reasonable system parameters.

\end{document}